\documentclass[twocolumn]{revtex4-1}          % twocolumn
\usepackage{graphicx, color}
\usepackage{amsmath, amssymb, amsfonts, mathrsfs}
\usepackage{times}
\usepackage{latexsym, color, amssymb}
\begin{document}
\title{Rheological properties of liquids under conditions of elastohydrodynamic lubrication}
\author{Vikram Jadhao}
\email{vjadhao@iu.edu}
\affiliation{Intelligent Systems Engineering, Indiana University, Bloomington, Indiana 47408}
\author{Mark O. Robbins}
\email{mr@jhu.edu}
\affiliation{Department of Physics and Astronomy, Johns Hopkins University, Baltimore, Maryland 21218}

\begin{abstract}
	There is an ongoing debate concerning the best rheological model for liquid flows in elastohydrodynamic lubrication (EHL). Due to the small contact area and high relative velocities of bounding solids, the lubricant experiences pressures in excess of 500 MPa and strain rates that are typically $10^5 -10^7$ $\textrm{s}^{-1}$. The high pressures lead to a dramatic rise in Newtonian viscosity $\eta_{N}$ and the high rates lead to large shear stresses and pronounced shear-thinning. This paper presents detailed simulations of a model EHL fluid, squalane, using nonequilibrium molecular dynamics methods to extract the scaling of its viscosity with shear rate ($10^5 - 10^{10}$ $\textrm{s}^{-1}$) over a wide range of pressure $P$ (0.1 MPa to 1.2 GPa), and temperature $T$ ($150 - 373$ K). Simulation results are consistent with a broad range of equilibrium and nonequilibrium experiments. At high $T$ and low $P$, where $\eta_{N}$ is low, the response can be fit to a power-law, as in the common Carreau model. Shear-thinning becomes steeper as $\eta_{N}$ increases, and for $\eta_{N}\gtrsim 1$ Pa-s, shear-thinning is consistent with the thermally activated flow assumed by another common model, Eyring theory.
	Simulations for a bi-disperse Lennard-Jones (LJ) system show that the transition from Carreau to Eyring is generic. For both squalane and the LJ system, the viscosity decreases by only about a decade in the Carreau regime, but may fall by many orders of magnitude in the Eyring regime. Shear thinning is often assumed to reflect changing molecular alignment, but the alignment of squalane molecules saturates after the viscosity has dropped by only about a factor of three. In contrast, thermal activation describes shear thinning by six or more decades in viscosity. Changes in the diagonal elements of the stress tensor with rate and shear stress are also studied. 
\end{abstract}

\maketitle

\section{Introduction}
Lubrication is commonly used to reduce friction and wear in mechanical devices.
In demanding applications,
lubricated machine components with nonconforming contacting elements, such as gears, rolling bearings, and cam followers,
often operate in the elastohydrodynamic lubrication (EHL) regime.
In this limit, high normal loads and small contact areas compress the
lubricant to pressures in excess of 500 MPa,
increasing the Newtonian viscosity $\eta_N$ by orders of magnitude.
At the same time, the high relative velocities of solid components
lead to strain rates $\dot{\gamma}$ that typically exceed $10^5 \, \textrm{s}^{-1}$.
The high $\eta_N$ helps to prevent squeeze out and maintain a thicker lubricant film,
while substantial shear thinning at the high strain rates limits
the rise in shear stress.
Both effects lower friction and can limit wear.

Accurate models describing the rheological properties of liquids under EHL lubrication are critical to predicting and optimizing machine performance, but have proved challenging to develop.
One reason is that it is difficult to reach the relevant pressures and strain rates in laboratory experiments.
Specially designed rheometers have been able to exceed 1 GPa at $10^4 \, \textrm{s}^{-1}$ and reach shear stresses of up to approximately 30 MPa \cite{bair.proc.inst.mech.2002,bair.tbl.2006,Bair2007,spikes.tbl.2014}, but may be affected by unmeasured temperature increases during shear at high rates \cite{spikes.tbl.2014}.
Experiments on idealized EHL contacts can go to the higher rates, pressures and shear stresses ($\sim 150$ MPa) important in real devices \cite{spikes.tbl.2014,dini.spikes.2017}
but measure an average stress over contact regions with a range of local strain rates,
pressures and temperatures \cite{bair.tbl.2015,Spikes2015}.
A second reason is that EHL conditions present fundamental theoretical challenges. 
They are in the regime where fluids exhibit poorly understood glassy rheology,
with rapid rises in $\eta_N$ with pressure, strongly nonlinear viscoelastic
behavior, rapid shear thinning, and changes in molecular order.
While there is a general agreement on the qualitative changes
in shear stress $\sigma$, the underlying mechanisms of shear thinning and
the functional form of the rheological response are hotly debated \cite{spikes.tbl.2014,bair.tbl.2015,Spikes2015}.

Two phenomenological models at the center of this active debate are based on work by Eyring \cite{eyring.jcp.1936,ewell.eyring.jcp.1937} and Carreau \cite{carreau.1972}, respectively.
The Eyring model assumes that flow occurs by thermal activation over
a single characteristic energy barrier height and
predicts that $\sigma$ rises as log($\dot{\gamma}$) at high $\dot{\gamma}$.
In the Carreau model, $\sigma$ increases as a power law, $\sigma \propto \dot{\gamma}^n$, with $n$ typically
around 0.5.
As discussed below, this power law behavior can arise from several different
physical pictures of flow,
including thermal activation over a broad range of energy barrier heights.

Both of these rheological models can be fit to existing rheometer data on simple fluids that extends up to rates of about $10^4$ $\textrm{s}^{-1}$.
The two models predict dramatically different shear stresses and viscosities when extrapolated to rates ($\dot{\gamma}> 10^5$ $\textrm{s}^{-1}$) that are important in EHL,
but inaccessible to rheometers \cite{spikes.tbl.2014}.
Resolving the debate about behavior at these rates is of great practical and fundamental interest.
In practical terms, it will yield accurate quantitative information about the macroscopic flow properties that is essential in predicting EHL friction and guiding optimal synthetic lubricant design. From a fundamental perspective, it will advance our understanding of the underlying microscopic (molecular-level) mechanisms of shear-thinning, changes in viscosity near the glass transition, and
the nature of flow in glassy systems.

This paper presents detailed nonequilibrium simulations of rheological behavior under EHL conditions that can distinguish between the above models and make contact to existing experiments on the frequently studied model lubricant squalane \cite{bair.proc.inst.mech.2002,bair.tbl.2006,spikes.tbl.2014,dini.spikes.2017,bair.cummings.prl.2002,mg.jcp.1995,moore.aiche.1997,moore.jcp.2000,harris2009temperature,schmidt.2015,bair2018new}.
A previous paper on the nature of the glass transition used some of the results
shown below to determine the Newtonian viscosity of squalane and obtained good agreement with experiment over a wide range of $P$ and $T$ \cite{jadhao.robbins.2017}.
Here, the nature of shear thinning is studied by evaluating the
shear stress, viscosity, normal stress, and molecular order for
a wide range of strain rates ($10^{5}$ -- $10^{10}$ $\textrm{s}^{-1}$) and thermodynamic conditions (pressures: 0.1 MPa -- 1.2 GPa; temperatures: 150 -- 373 K).
At high temperatures and low pressures, where $\eta_N$ is low,
shear thinning can be described by a power-law, as in the Carreau model.
As $\eta_N$ rises above $\sim 1$ Pa-s, squalane begins to exhibit glassy flow and shear-thinning is increasingly dominated by thermally-activated rearrangements. Results from this high $\eta_N$ regime can be collapsed onto Eyring theory over almost 30 orders of magnitude in rate using time-temperature superposition.
To illustrate that this transition from Carreau to Eyring behavior as $\eta_N$ increases is generic,
data for a commonly studied model of bi-disperse Lennard-Jones spheres is also
presented.
We argue that many energy barriers may contribute to motion at low $\eta_N$,
leading to Carreau-like behavior.
As $\eta_N$ increases, the response is increasingly dominated by thermally activated rearrangements over the lowest energy barrier and thus can be described with Eyring theory.

Shear thinning is often attributed to molecular ordering.
While squalane molecules tend to align along the flow direction,
the degree of order saturates before the viscosity has dropped by more than
half a decade.
In contrast, Eyring theory can describe shear thinning by more than six decades in viscosity and over more than eight decades in rate.

The simulations performed here are at constant density and lead to changes in the diagonal components of the pressure tensor.
These changes are strongly correlated with the shear stress in both Eyring and Carreau regimes.
The components along the velocity gradient and in the vorticity direction
increase nearly quadratically with shear stress over the entire range of rates.
In contrast, the component along the flow direction first drops and then rises more slowly than other components.
The differences between diagonal components, or normal stresses, show a logarithmic rise with rate at high $\dot{\gamma}$.
When plotted against shear stress, the normal stresses
	for systems at different $T$ and $P$
collapse well onto universal curves in the shear-thinning (non-Newtonian) regime.
Moreover, systems in both the Eyring and Carreau shear-thinning regimes
collapse onto the same curve.
The ratio of shear stress to first normal stress approaches a constant, analogous to a friction coefficient, at high rates.
Experiments are often performed at a nearly constant value of a component of the pressure tensor and the implications of associated changes in density for variations in shear stress with rate are discussed.

The next section provides a brief overview of the rheological models that simulations are compared to.
Section \ref{sec:method} presents details of the interaction potentials and simulation methods used.
Results for shear stress, viscosity, molecular alignment, and components of the pressure tensor are given in Sec. \ref{sec:results}.
The final section presents a discussion and conclusions.

\section{Phenomenological Models for Non-Newtonian Flow}\label{sec:models}
At sufficiently low strain rates, fluids are near their equilibrium state and have a constant Newtonian viscosity $\eta_{N}$ that is consistent with the fluctuation-dissipation relation.
Higher shear rates lead to changes from equilibrium correlation functions and a strain-rate-dependent viscosity $\eta (\dot{\gamma})$.
In most cases, changes in molecular order make flow easier, and $\eta$ decreases with increasing rate.
This shear-thinning behavior is important in EHL contacts because it lowers the frictional stress $\sigma = \dot{\gamma} \eta(\dot{\gamma})$ at high sliding velocities.

In the following subsections, we briefly review two classes of phenomenological models that are commonly used to describe shear thinning in EHL and other non-Newtonian flows \cite{spikes.tbl.2014}.
The first is the Eyring theory of thermally activated flow over a single characteristic energy barrier height. It yields a logarithmic rise in stress with $\dot{\gamma}$ at high rates. The second class of models can be derived by assuming a broad distribution of activation barrier heights and leads
to a power law scaling of stress with rate.

In Sec. \ref{sec:results} we show that both types of models describe simulations of simple fluids,
but in different regimes of temperature and pressure.
Power law behavior is prevalent at low $\eta_{N}$ where thermal energies are large compared to energy barriers so that many barriers may contribute.
At large $\eta_{N}$, activation is rare, the lowest barrier dominates, and the Eyring model provides a better description.

\subsection{Eyring model}

The Eyring model is based on the idea that shear flow is a stress-biased thermally activated process \cite{eyring.jcp.1936,ewell.eyring.jcp.1937}.
The model assumes that shear can only occur through molecular rearrangements that require activation over potential energy barriers with a single characteristic height.
In equilibrium, the barrier height has the same value $E$ for molecular rearrangements that allow shear in either direction and there is no net flow.
When a shear stress $\sigma$ is applied,
the model assumes that $E$ is lowered in the direction of applied stress (forward) and raised in the reverse direction (backward) by an amount linearly proportional to $\sigma$.
The constant of proportionality $V^*$ has dimensions of volume and may depend on temperature. While it is called the ``activation volume,'' it represents the sensitivity of the energy barrier height to stress rather than a literal volume of fluid involved in rearrangements.

Given the above assumptions, the rates of rearrangements in the forward and
backward directions are
$\nu_f=\nu_0 e^{-(E - \sigma V^*)/k_{B} T}$ and $\nu_b=\nu_0 e^{-(E + \sigma V^*)/k_{B} T}$, respectively, where $k_{B}$ is the Boltzmann constant and $\nu_0$ is the frequency of attempted rearrangements.
Assuming the strain rate $\dot{\gamma}$ is proportional to the excess of forward rearrangements, one obtains:
\begin{equation}
\label{eq:eyring.rate.stress}
\dot{\gamma} = c (\nu_f - \nu_b) =2c\nu_0e^{-E/k_{B} T} \sinh(\sigma / \sigma_E),
\end{equation}
where $\sigma_E = k_{B} T / V^*$ is a parameter related to the activation volume and is called the Eyring stress. The unknown quantities $c$ and $\nu_0$ can be eliminated by equating the linear response at low rates to the Newtonian viscosity $\eta_{N}$.
Expanding the sinh in Eq. \ref{eq:eyring.rate.stress} for small $\sigma/\sigma_E$, one finds
\begin{equation}\label{eq:newtonian.viscosity}
	\eta_{N} = \frac{\sigma_E}{2c\nu_0}e^{E/k_{B} T}
\end{equation}
and solving for $\sigma$ in Eq.~\ref{eq:eyring.rate.stress}, one obtains
\begin{equation}\label{eq:eyring}
	\sigma = \sigma_E \, \sinh^{-1}(\eta_{N} \dot{\gamma}/\sigma_E),
\end{equation}
where $\textrm{sinh}^{-1}$ is the inverse hyperbolic sine function.

Equation \ref{eq:eyring} is the Eyring equation that we will compare to simulations and experiments.
It predicts Newtonian behavior with a constant viscosity for
$\sigma \ll \sigma_E$ and $\dot{\gamma} \ll \dot{\gamma}_{E}$, where $\dot{\gamma}_{E}=\sigma_E/\eta_{N}$
is called the Eyring rate.
At larger rates and stresses, the stress rises sublinearly, implying shear thinning with
\begin{equation}
	\frac{\eta}{ \eta_{N}} = \frac{\dot{\gamma}_{E}}{\dot{\gamma}} \sinh^{-1}(\dot{\gamma} /\dot{\gamma}_{E}).
\label{eq:eyring.viscosity.rate}
\end{equation}
Limiting behavior for high rates and stresses sets
in for $\dot{\gamma} \gg \dot{\gamma}_{E}$ and $ \sigma \gg \sigma_E$.
The stress then rises logarithmically with rate, 
\begin{equation}
	\frac{\sigma }{ \sigma_E} = \log(2 \dot{\gamma}/\dot{\gamma}_{E})
\end{equation}
and
\begin{equation}
	\frac{\eta }{ \eta_{N}}= \frac{\dot{\gamma}_{E}}{\dot{\gamma}} \log(2 \dot{\gamma}/\dot{\gamma}_{E}).
\label{eq:highrate.eyring.viscosity.rate}
\end{equation} 

The assumption of a single energy barrier is clearly unrealistic.
The Eyring model can be generalized to include an arbitrary distribution of barriers and activation volumes \cite{ree1955theory}.
Of course this introduces a large number of parameters and reduces the predictive power of the model.
A greater difficulty is the assumption that molecular motion can be described as discrete, rare transitions between local energy minima.
This picture breaks down in low viscosity fluids where molecules are in almost continual motion rather
than hopping between cages formed by their neighbors.
The next class of models is better suited to this limit.

\subsection{Power-law fluid models}

An alternate starting point for describing shear flow is to assume that there is a very broad
distribution of barrier heights and associated time scales.
This idea has been implemented in a range of different approaches,
including shear transformation zone theory \cite{falk.langer.1998}, the soft glassy rheology model \cite{sollich.cates.1997}, and molecular network theory and the associated Carreau equation \cite{carreau.1972}.
The common prediction is that stress rises with rate as a power law at high rates:
\begin{equation}
	\sigma = B \dot{\gamma}^{n}
\end{equation}
where $B$ is a constant and the exponent $n$ is between $0$ and $1$.
The corresponding viscosity shows power-law shear thinning as a function of rate:
\begin{equation}
	\eta = B \dot{\gamma}^{n-1}. \label{eq:carreau.viscosity.rate}
\end{equation}
Following many studies of EHL,
we will fit data to the Carreau equation
\begin{equation}
	\eta = \eta_{N} \left[ 1 + \left(\dot{\gamma} / \dot{\gamma}_0 \right)^2  \right]^{\frac{n-1}{2}} 
\label{eq:carreau.viscosity.rate2}
\end{equation}
which has a specific form of the crossover from Newtonian behavior at $\dot{\gamma}$ below a characteristic rate $\dot{\gamma}_0$ to power law behavior at $\dot{\gamma} \gg \dot{\gamma}_0$.

Power-law shear-thinning can also result from fluid models where a continuous increase in some type of order with increasing rate leads to reductions in dissipation during flow.
For example, individual molecules may stretch or align and groups of molecules may order into lines to reduce friction.
In most cases, the degree of order that can be induced is finite and thus the range of power law scaling will be limited.
For example a polymer chain can go from an equilibrium random walk to a perfectly straight chain, but then
shear thinning due to changing order must saturate.
Because lubricant molecules are typically short, the degree of change in alignment is limited.

\begin{figure}[t]
\centerline{\includegraphics[scale=0.3]{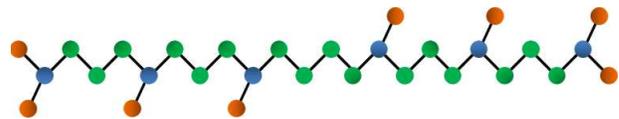}}
\caption{Sketch of a squalane molecule showing carbons and hydrogens combined into united-atoms (circles) that are connected by covalent bonds (lines). Red, green, and blue circles indicate carbon atoms with three, two, and one hydrogen atoms, respectively.}
\label{fig:squalane}
\end{figure}

\section{Models and Simulation Methods}\label{sec:method}
To explore the range of possible fluid behavior we study both a chemically specific model for squalane (Fig. \ref{fig:squalane}) and a generic model for small molecule glass-formers.
We first describe general simulation methods.
Model potentials and specific simulation details
are provided in the following subsections.

In all cases,
simulations were performed using the parallel molecular dynamics package LAMMPS from Sandia National Laboratories \cite{plimpton.lammps}.
Periodic boundary conditions were used to prevent edge effects and
temperature was maintained with a Nos\'e-Hoover thermostat.
Initial states were prepared at a temperature $T$ and density $\rho$ where diffusion was rapid and equilibrium was reached rapidly.
Then $T$ and $\rho$ were adjusted gradually to the desired values.
The rate of changes in $T$ and $\rho$ was decreased as diffusion slowed,
but it was not possible to ensure equilibrium was reached at state points deep
in the glassy regime.
Similar problems arise eventually in any experimental study.
While this leads to uncertainty in the equilibrium pressure associated with the values of $\rho$ that we study,
our nonequilibrium simulations rapidly approach the correct steady state response (stresses, alignment, etc.) for
the chosen density and are not sensitive to any deviation from equilibrium structure in the state before shear is applied.

To study steady-state flow behavior we impose simple shear (planar Couette flow)
using the SLLOD equations of motion \cite{evans.morris}.
The periodic box was sheared along the $x$-axis at a constant rate to impose the desired average strain rate $\dot{\gamma}=\delta v_x/\delta y$, with the velocity
$v_x$ along the $x$-axis and velocity gradient along the $y$-axis.
Within statistical fluctuations,
the velocity profiles were consistent with a constant local shear rate.
There was no evidence of shear localization
and thus measured quantities represent the response to homogeneous flow.

For each shear rate $\dot{\gamma}$, the system was sheared until it reached steady state, as indicated by the saturation of the shear stress $\sigma$ and other statistical quantities.
While the time to reach equilibrium rises beyond simulation time scales in the glassy regime, steady state for a given $T$ and $\rho$ is always achieved in a time of order $1/\dot{\gamma}$ because shearing the system forces molecules to sample new configurations.
The total strain required to reach steady state was of order 5 to 10 far from the glass transition and up to $100$ in cases where shear produced large changes in molecular order. In such cases, equilibration can be accelerated by gradually changing $\dot{\gamma}$.
After reaching steady state, simulations were run for strains large enough (typically $>5$; $>30$ for $\dot{\gamma} \gtrsim 10^7$) to achieve the desired statistical accuracy in average quantities such as the stress and viscosity $\eta \equiv \sigma/\dot{\gamma}$.

Nonequilibrium molecular dynamics simulations usually make the assumption that systems are not too far from a local thermal equilibrium.
For example, the temperature is calculated from the thermal kinetic energy associated with peculiar velocities - the velocities relative to the mean flow.
The equations of motion are then modified to maintain the desired value of thermal kinetic energy.
This approach breaks down, and temperature is hard to define, when changes in velocity on the scale $a$ of the distance between neighboring molecules becomes comparable to the thermal velocity $v_T$ of molecules.
Studies of simple systems suggest that these effects are small for \cite{thompson90,khare1997rheological,loose89} 
\begin{equation}
	\dot{\gamma} < \dot{\gamma}_{max} = 0.1 v_T / a.
\label{eq:ratemax}
\end{equation}
Experiments are typically performed at much smaller rates.

Other characteristic rates can also lead to new behavior and deviations
from Eyring or Carreau fits.
For example, new dissipation mechanisms are possible as the
rate approaches characteristic vibrational frequencies or structural relaxation times.
Intramolecular vibrations in squalane are $10^{12} \ \textrm{s}^{-1}$ and
above. Sound waves with wavelengths on the scale of the simulation box have frequencies $\sim 3 \times 10^{11} \ \textrm{s}^{-1}$.
Studies of glassy Lennard-Jones systems indicate that disorder may
lead to even longer relaxation times \cite{rr.pre}.
Such effects are discussed further below.

\subsection{Squalane}

Squalane or 2,6,10,15,19,23-hexamethyltetracosane ($\textrm{C}_{30}\textrm{H}_{62}$) is a branched alkane with a $\textrm{C}_{24}$ backbone and six methyl side groups.
As shown in Fig. \ref{fig:squalane}, the side groups are placed symmetrically about the backbone in a way that frustrates crystallization.
Squalane has been widely studied as a model experimental EHL system because it is
an excellent glass former, is representative of low-molecular-weight lubricant base oils, and is available in almost pure form at relatively low cost \cite{bair.cummings.prl.2002,bair.tbl.2006,spikes.tbl.2014,dini.spikes.2017,schmidt.2015}.
Due to its relatively simple chemical structure and small molecular size (end-to-end distance $\sim 2$ nm), squalane has also been chosen for several computational studies of shear-thinning \cite{dini.spikes.2017,bair.cummings.prl.2002,moore.aiche.1997,moore.jcp.2000,jadhao.robbins.2017}.
Early work was limited to short simulation times and examined very high rate ($\sim 10^8 - 10^{11} \ \textrm{s}^{-1}$) behavior at state points where the Newtonian viscosity was low.
Increased computer power has enabled recent flow studies under pressure
and temperature conditions prevalent in EHL \cite{dini.spikes.2017,jadhao.robbins.2017}, and at rates as low as $10^5 \ \textrm{s}^{-1}$ \cite{jadhao.robbins.2017}.

We model squalane using a united-atom (UA) potential based on the model developed by Mondello and Grest (MG) for branched alkanes \cite{mg.jcp.1995,mg.jcp.1996}.
Following other simulations of branched alkanes, including squalane \cite{moore.jcp.2000,moore.fluid.mech.2000}, the fixed bond lengths of the MG model
are replaced by stiff harmonic bond-stretching potentials.
In the UA description, hydrogens are lumped together with carbons into UA groups such as $\textrm{CH}_3$, $\textrm{CH}_2$, and CH. As shown in Fig.~\ref{fig:squalane}, each squalane molecule is represented by 30 UAs. Within each molecule, the UAs interact via a harmonic bond-bending term, a harmonic bond-stretching term, a torsional potential characterizing the rotational barrier around nonterminal bonds, and a harmonic bending term to prevent umbrella ($sp_3$) inversion at tertiary carbon branch points. There is an additional nonbonded Lennard-Jones potential between united atoms on different molecules or separated by at least
four bonds on the same molecule.
We refer to the original papers \cite{mg.jcp.1995,moore.jcp.2000,moore.fluid.mech.2000} for further details about the model and the potential parameters.
A LAMMPS file with the potential parameters is provided in the supplemental
data of Ref. \cite{jadhao.robbins.2017}.

\begin{figure}[t]
\centerline{\includegraphics[scale=0.32]{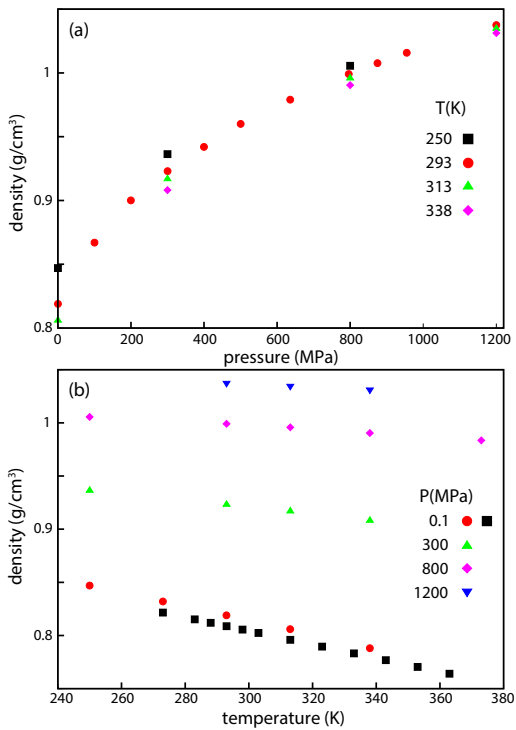}}
\caption{(a) Density of squalane as a function of pressure from equilibrium 
simulations at the indicated temperatures.
(b) Density of squalane as a function of temperature from simulations (colored symbols)
at the indicated pressures and experimental data \cite{harris2009temperature} (black squares) at ambient pressure.}
\label{fig:density}
\end{figure}

\begin{figure}[t]
\centerline{\includegraphics[scale=0.32]{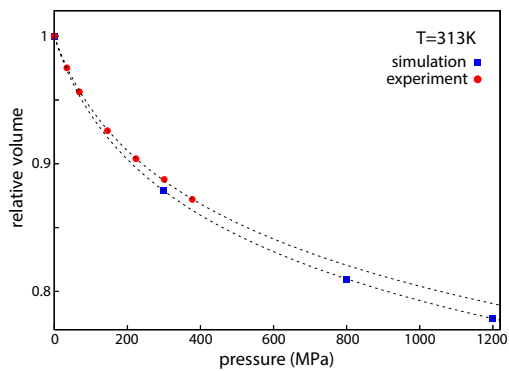}}
\caption{Relative volume of squalane compared to ambient pressure data at 313 K from
simulations and experiments \cite{bair.tbl.2006} (points) and fits to the Tait equation (lines).
}
\label{relative.volume.pressure}
\end{figure}

Near room temperature, the thermal velocity of squalane is of order $100$ m/s and, at the pressures we consider, the spacing between molecules is of order $1$ nm. From Eq. \ref{eq:ratemax}, this implies $\dot{\gamma}_{max} \sim 10^{10} \ \textrm{s}^{-1}$ and we will only report simulations up to this range.
Most of our simulations of squalane used $N_m=125$ molecules ($N=3750$ UAs) in a periodic unit cell with cubic geometry.
As noted in Ref. \cite{jadhao.robbins.2017}, simulations with 8 times as many UAs and with an all-atom potential, gave statistically similar results.
An initial state was created at
room temperature $T=293$ K and ambient pressure $P=0.1$ MPa where
squalane is a simple low-viscosity liquid.
High-pressure states were generated by gradually compressing this state at
constant $T$ until the density or hydrostatic pressure $P$ reached the desired value.
For a given $P$, liquid configurations at lower (higher) temperatures were generated by quenching (heating) the room temperature configuration at constant volume. 
The system was then allowed to equilibrate under constant volume at the desired temperature.
The density was then adjusted at the final temperature using a constant NPT simulation.
Typical equilibration times were $\approx 100$ nanoseconds and the simulation time step was 1 femtosecond.

The variation of $\rho$ with $P$ and $T$ is shown in Fig. \ref{fig:density}.
We focus on the limited range of $T$ where there is experimental data \cite{harris2009temperature,fandino2005compressed}
at ambient pressure (Fig. \ref{fig:density}(b)).
The MG model slightly over-estimates the density by about one percent
at ambient pressure.
The offset from experimental data is nearly constant, so the coefficients of 
thermal expansion are consistent within our statistical accuracy.
Figure \ref{relative.volume.pressure} compares the fractional change
in volume per molecule with pressure from experiments and simulations at $T=313$ K.
Both experiment and simulation are well fit by the Tait equation \cite{bair.tbl.2006},
which is commonly used to interpolate and extrapolate experimental results.
The simulation density increases slightly more rapidly with pressure.

As $T$ decreases and $\rho$ increases, the relaxation time grows, and eventually simulations and experiments fall out of equilibrium.
The onset of this behavior occurs later in experiments because of their longer time scales and the implications for comparison between simulations and experiments are discussed further below.
We will label results for a given state point by the target pressure used to equilibrate the initial state.
Extrapolations in Section \ref{sec:stresstensor} suggest this may deviate by up to 3 to 5\% from the equilibrium
pressure in systems with the highest $\eta_N$.

\subsection{Bi-disperse Lennard-Jones Fluid}
\label{sec:LJglass}

One of the most commonly used systems in simulation studies of the glass transition is a bi-disperse mixture of particles interacting with the truncated Lennard-Jones (LJ) potential.
Particles of type $I$ and $J$ separated by a distance $r$ interact with the potential
\begin{equation}
	U_{IJ}(r)=4 u_{IJ} \left[ \left( \frac{a_{IJ}}{r} \right)^{12} - 
	\left( \frac{a_{IJ}}{r} \right)^6 \right] \ , \ {\rm for } \ \ r < 1.5 a_{IJ}.
\end{equation}
$U_{IJ}$ is zero for $r > 1.5 a_{IJ}$, and
$u_{IJ}$ and $a_{IJ}$ characterize, respectively, the binding energy and diameter for the corresponding particle types.
Crystallization is suppressed by choosing a ratio of particle sizes that frustrates packing, and using interaction strengths that prevent demixing.
As in Ref. \cite{rr.pre}, we used $a_{AA}=a$, $a_{AB}=0.88 a$, $a_{BB}=0.8 a$, $u_{AA}=u$, $u_{AB}=1.5u$, and $u_{BB}=0.5u$, and will express results in terms of $u$ and $a$.
Time is expressed in terms of $\tau_{LJ} \equiv \sqrt{ma^2/u}$, where
all particles have mass $m$. 

Simulations used a time step of $0.005 \tau_{LJ}$ and a Nos\'e-Hoover thermostat with time constant $\tau_{LJ}$. Equilibrium states were made at temperatures $k_B T/u$ between 0.24 and 0.3,
where past studies showed rapid changes in viscosity \cite{rr.pre}.
A disordered high temperature state was slowly quenched to the desired temperature while
maintaining zero pressure with a Nos\'e-Hoover barostat.
The density was then fixed for simulations of shear.
The equilibrium pressure at the chosen densities deviated from zero by less than
$0.1 u/a^3$.
Simulations are only shown for $\dot{\gamma} \leq 10^{-2} \tau_{LJ}^{-1}$. The velocity difference between neighboring atoms at this rate is less than 1\% of the thermal velocity, but temperature rises were notable at higher rates and our focus is on isothermal behavior.

\section{Results}
\label{sec:results}

\subsection{Rate dependence of shear stress in squalane: Eyring vs. Carreau models}

\begin{figure}[t]
\centerline{\includegraphics[scale=0.37]{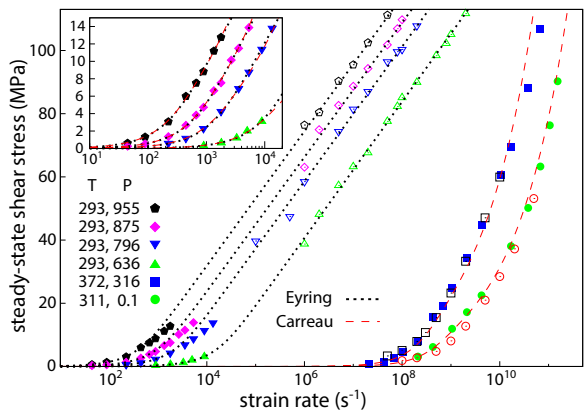}}
\caption{The rate dependent steady-state shear stress $\sigma$
	(solid symbols) from previous experiments \cite{bair.proc.inst.mech.2002,bair.cummings.prl.2002} and simulations \cite{bair.cummings.prl.2002} at the indicated $T$ in K and $P$ in MPa. Open symbols show our simulations at the experimental state points and
	state points close enough to past simulations, 373 K and 300 MPa (squares) and 313 K and 0.1 MPa (circles), that deviations are comparable to statistical fluctuations. These are comparable to the symbol size for our data.
	Dotted black lines show Eyring fits with $\sigma_E \approx 9.3$ MPa and dashed red lines show Carreau fits with $n\approx 0.4$.
	The inset shows a blow up of the experimental data, and Eyring fits with $\sigma_E \approx 5.5$ MPa and Carreau fits with $n \approx 0.45$.
}
\label{eyring.carreau}
\end{figure}

Squalane has played an important role in the debate about rheological models of shear-thinning  \cite{spikes.tbl.2014,dini.spikes.2017,bair.tbl.2015,Spikes2015,ewen2018advances}.
Figure \ref{eyring.carreau} reproduces early results (solid symbols) for the stress from simulations and experiments 
\cite{bair.proc.inst.mech.2002,bair.cummings.prl.2002}.
As shown in the inset, experimental data are equally well fit by Carreau or Eyring equations over the limited range of rates.
Although they were at very different $T$ and $P$, simulation results appeared to have a similar form and covered a wider range of rates.

A common assumption is that changing $T$ and $P$ does not change the functional form of shear thinning, but merely scales the time for all relaxation
processes by a constant factor. When this time-temperature superposition holds, plots of $\eta/\eta_{N}$ collapse when $\dot{\gamma}$ is multiplied by a single scale factor $a_r(P,T)$.
The data from Refs. \cite{bair.proc.inst.mech.2002,bair.cummings.prl.2002} (solid symbols in Fig. \ref{eyring.carreau}) appeared consistent with this scaling and indicated that the viscosity of squalane followed a Carreau law with $n=0.463$.
However, the fit was over a limited range, with the viscosity decreasing by only a factor of 3 for experiment and 10 for simulations.

To test the proposed time-temperature scaling, we performed constant density simulations near the different state points used in experiments and simulations.
As shown in Fig. \ref{eyring.carreau}, our results (open symbols) are consistent with past
data and, by extending the range of rates, reveal that
previous simulations and experiments were in different regimes.
For the low viscosity states in past simulations (squares and circles), the viscosity and stress are well fit by the Carreau model with $n \sim 0.4$.
For the high viscosity regime of experiments, our simulations and experiments are consistent with the Eyring equation over more than six orders of magnitude in strain rate.
Any Carreau fit to the experimental data is orders of magnitude higher than the simulation results.

\begin{figure}[t]
\centerline{\includegraphics[scale=0.37]{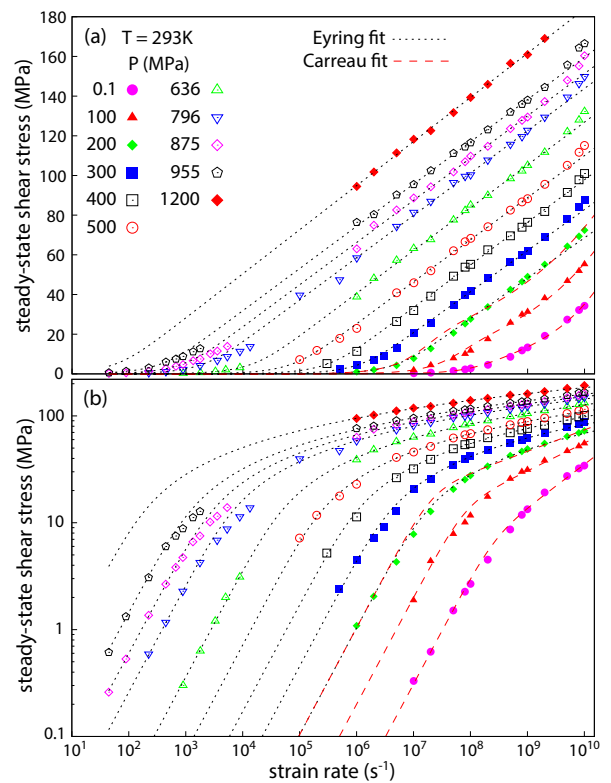}}
\caption{(a) Steady-state shear stress $\sigma$ of squalane as a function of strain rate $\dot{\gamma}$ at $T=293$ K for pressures in the range $P = 0.1 - 1200$ MPa. Symbols for $\dot{\gamma} \ge 10^5$ $\textrm{s}^{-1}$ correspond to simulation results, and symbols at lower rates are experimental values \cite{bair.proc.inst.mech.2002,bair.cummings.prl.2002}. Black dotted lines are Eyring model fits to simulation data and red dashed lines are Carreau fits. Error bars are comparable to or smaller than the symbol size.
(b) Same data plotted on a log-log plot. Fits are constrained to the limiting Newtonian viscosity from simulations for $P \leq 500$ MPa and from nonequilibrium experiments for $P$ between $636$ MPa and $955$ MPa. The Newtonian viscosity is fit at $1.2$ GPa since the experimental value is not available.
}
\label{atT293}
\end{figure}

\begin{figure}[t] 
\centerline{\includegraphics[scale=0.37]{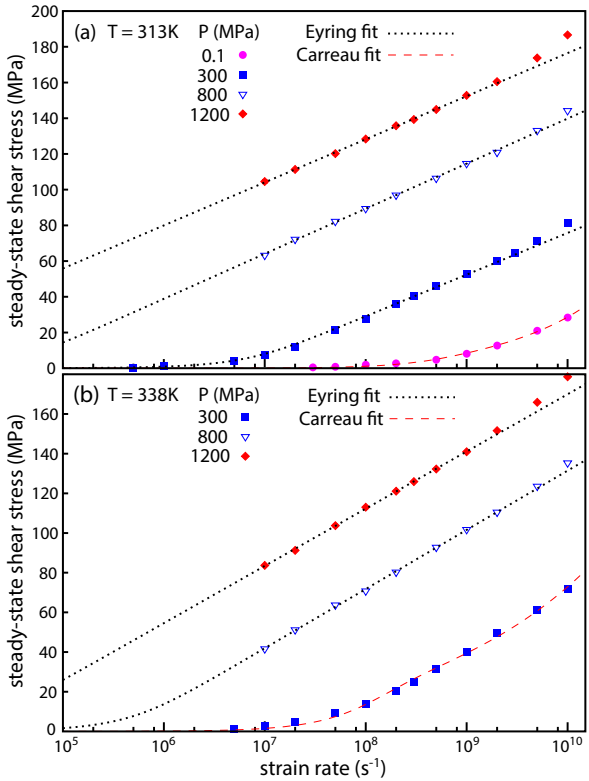}}
\caption{Steady-state shear stress as a function of strain rate from simulations of squalane at (a) 313 K and (b) 338 K and the indicated pressures.
Black dotted lines are fits to the Eyring equation and red dashed lines are fits to the Carreau equation.
}
\label{atT313}
\end{figure}

To examine the crossover between Carreau and Eyring regimes with increasing $\eta_{N}$, we performed simulations at $T=293$ K for pressures between $0.1$ MPa and $1200$ MPa, focusing on values where there is experimental data \cite{bair.proc.inst.mech.2002,bair.tbl.2006,bair.cummings.prl.2002}.
The results are shown in Fig. \ref{atT293} along with fits to the Eyring and Carreau equations.
For $P \leq 500$ MPa, simulations reached the Newtonian regime and the limiting $\eta_N$ was used in the fits.
For $P$ between 636 MPa and 955 MPa, fits were constrained to have the experimental value of $\eta_N$.
No data were available to constrain $\eta_N$ at 1200 MPa, so $\eta_N$ was included in the fit.

As seen in the plot of stress vs logarithm of rate in Fig. \ref{atT293}(a), high-pressure results ($P \geq 300$ MPa) from experiments and simulations are consistent with the Eyring equation over up to 8 orders of magnitude in rate.
At high rates, $\sigma$ rises logarithmically with rate and the slope gives the Eyring stress $\sigma_E$, which is $\approx 9.3$ MPa for $500 \leq P \leq 955$ MPa, and $\approx 9.6$ MPa for 1200 MPa.
Fits to simulations extending to rates as low as $10^5 \ \textrm{s}^{-1}$ are consistent
with experiments at the same pressure that reach rates up to $10^4 \ \textrm{s}^{-1}$ and extend down into the Newtonian regime.
Although there is no nonequilibrium experimental data for $P \leq 500 $ MPa, simulations are consistent with the measured Newtonian viscosity \cite{bair.tbl.2006,jadhao.robbins.2017} (Fig. \ref{viscosity.rate.atT293K}).

As the pressure decreases, the range of rates where logarithmic behavior is observed decreases.   
The lower bound is set by $\dot{\gamma}_E=\sigma_E/\eta_{N}$ and, as noted in the previous section, temperature is hard to define at rates above $\dot{\gamma}_{max} \sim10^{10} \ \textrm{s}^{-1}$, which sets the upper bound.
For $P < 200$ MPa ($\eta_{N} \lesssim 1$ Pa-s), the plots in Fig. \ref{atT293}(a) show clear curvature and are not well fit by the Eyring equation. 
In contrast, this low pressure data is reasonably fit by the Carreau equation, which implies linear scaling in the log-log plot of Fig. \ref{atT293}(b).
Linear, Newtonian ($n=1$) behavior extends up to stresses of order 10 MPa and then the stress rises with a lower power law up to the highest rates and stresses (30 to 60 MPa).
There is a gradual decrease in the exponent $n$ of the power law as $P$ increases;
$n$ decreases from about 0.46 at 0.1 MPa to 0.2 at 200 MPa.

Figure \ref{atT313} shows that there are similar crossovers from Eyring to Carreau behavior with decreasing pressure at $T=313$ K and $338$ K.
The transition occurs at a higher pressure as $T$ increases, but always at a value of $\eta_{N}$ near 1 Pa-s.
For example, at $P=300$ MPa, $\eta_{N}$ drops from $0.9$ Pa-s to $0.15$ Pa-s as $T$ increases from
313 K to 338 K and the behavior changes from approximately Eyring to Carreau.

\begin{figure}[t] 
\centerline{\includegraphics[scale=0.37]{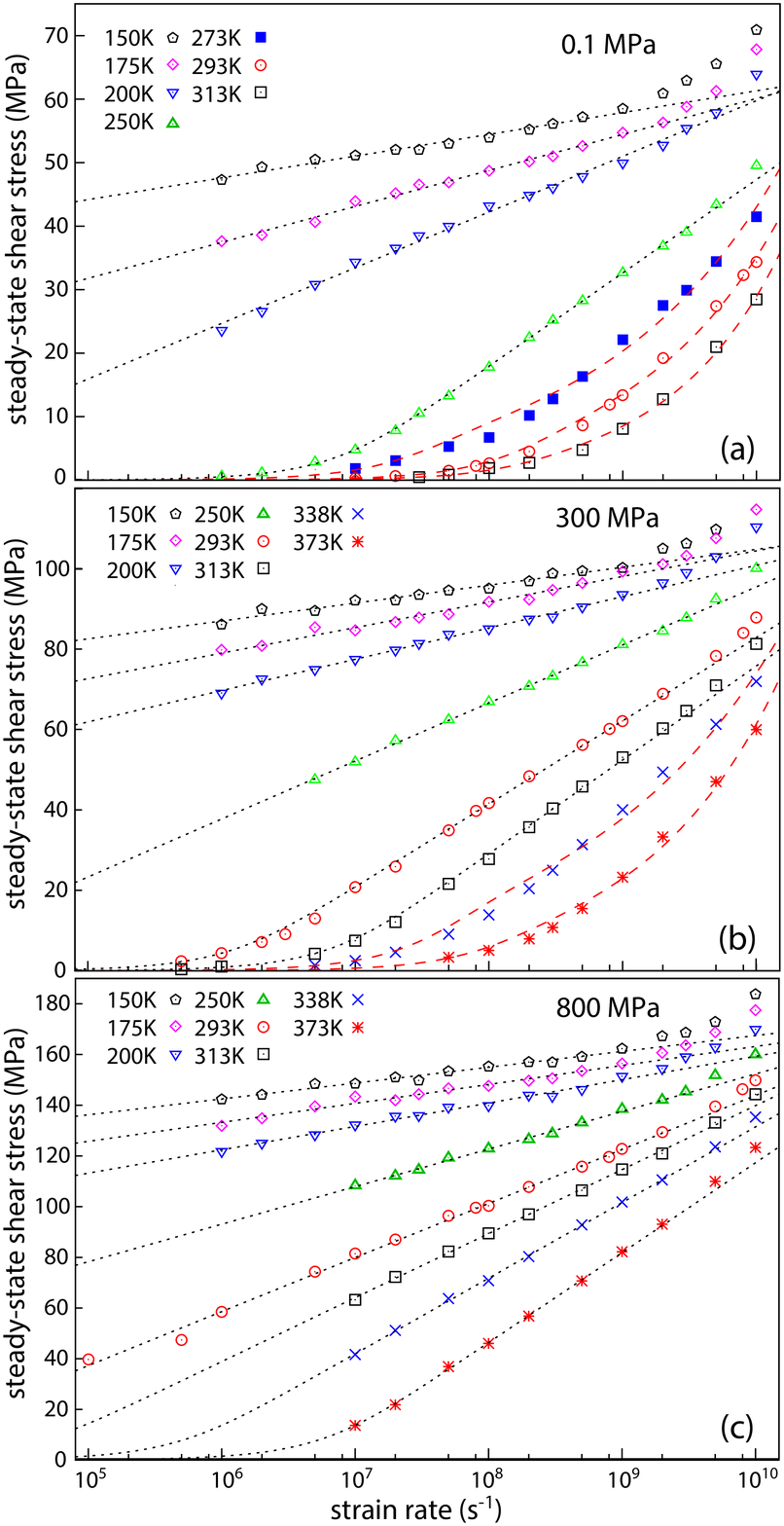}}
\caption{Steady-state shear stress as a function of strain rate from simulations of squalane at the indicated temperatures for (a) $P=0.1$ MPa, (b) $P=300$ MPa, and (c) $P=800$ MPa. Black dotted lines are Eyring fits for high viscosity states and red dashed lines are Carreau fits for low viscosity states.
}
\label{atP}
\end{figure}

The previous plots showed a transition from Eyring to Carreau behavior with increasing pressure.
Figure \ref{atP} shows results for the temperature dependence at fixed pressure.
At ambient pressure ($P=0.1$ MPa; Fig.~\ref{atP}(a)), Carreau behavior is observed for $T \gtrsim 273$ K.
As $P$ increases, Eyring behavior extends to higher temperatures and by $800$ MPa (Fig.~\ref{atP}(c)), all results are consistent with Eyring theory.
For all $T$ and $P$, we find that Eyring theory provides a reasonable fit
for systems with $\eta_{N} \gtrsim 1$ Pa-s while Carreau theory provides a better fit for $\eta_{N} \lesssim 1$ Pa-s.
Scaling collapses in the two regimes are shown in the next section.

\begin{figure}[t] 
\centerline{\includegraphics[scale=0.37]{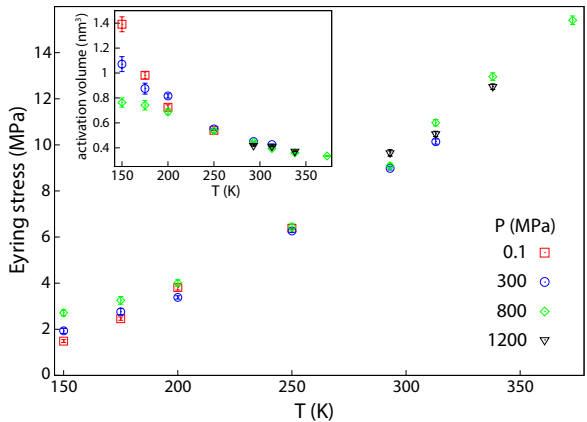}}
\caption{Eyring stress $\sigma_E$ as a function of temperature at the indicated pressures. Inset: Corresponding activation volume $V^*=k_BT/\sigma_E$ for the same $T$ and $P$.
}
\label{ActVol}
\end{figure}

The slope of each Eyring fit in Fig. \ref{atP} corresponds to the Eyring stress, $\sigma_E = k_B T/V^*$. Figure \ref{ActVol} shows the temperature dependence of $\sigma_E$ for different pressures.
In Eyring theory, $\eta/\eta_N =1/\sinh(1) \sim 85\%$ at $\sigma_E$, and shear thinning becomes rapid as $\sigma$ increases above $\sigma_E$.
Ref. \cite{spikes.tbl.2014} says that typical values for $\sigma_E$
under EHL conditions are between 5 and 10 MPa.
Our results for squalane at room temperature are at the high end of this range and $\sigma_E$ decreases steadily as $T$ is lowered.
The decrease is faster than would be predicted by a constant activation volume and the inset of Fig. \ref{ActVol} shows the inferred temperature dependence of $V^*$.
As noted above, $V^*$ represents the sensitivity of the activation barrier to stress rather than a literal volume of the material that moves.
Nonetheless, the high temperature values of $V^*$ correspond to roughly half the molecular volume of squalane, $0.8$ nm$^3$. The value of $V^*$ increases with decreasing temperature, rising by a factor of two to three by 150 K.
Little pressure dependence is evident at high $T$ but
the increase in $V^*$ appears to saturate at lower values for higher pressures.

\subsection{Shear thinning and time-temperature superposition in squalane }

\begin{figure}[t] 
\centerline{\includegraphics[scale=0.38]{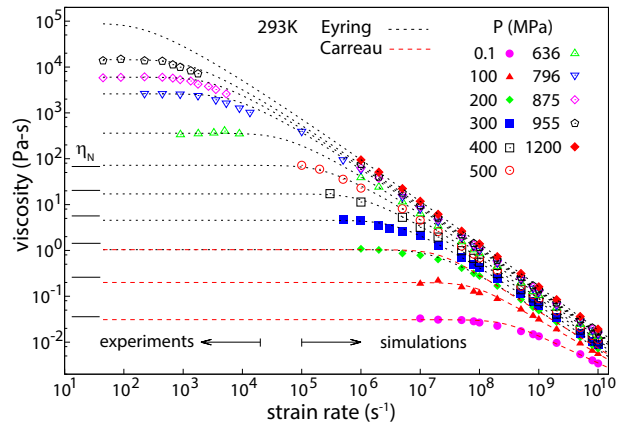}}
\caption{Steady-state viscosity $\eta$ of squalane as a function of strain rate $\dot{\gamma}$ at $T=293$ K for pressures in the range $P = 0.1 - 1200$ MPa. Symbols for rate $\dot{\gamma} \ge 10^5$ $\textrm{s}^{-1}$ correspond to simulation results, other symbols are experimental values \cite{bair.proc.inst.mech.2002,bair.cummings.prl.2002,bair.tbl.2006}. Black dotted lines are Eyring fits and red dashed lines are Carreau fits. Short black solid lines at low rates show the experimental value for the Newtonian viscosity $\eta_{N}$ at $P \leq500$ MPa \cite{bair.tbl.2006}.
}
\label{viscosity.rate.atT293K}
\end{figure}

While the traction force measured in EHL contacts scales with the shear stress,
rheologists often focus on plots of viscosity vs. rate.
Figure \ref{viscosity.rate.atT293K} shows the viscosities corresponding to the $T=293$ K data in Fig. \ref{atT293}.
Carreau theory predicts a power law decrease in viscosity, with
a slope of $-(1-n)$ in a log-log plot.
Data for $P < 300$ MPa in Fig. \ref{viscosity.rate.atT293K} are fit to the Carreau equation with values of $n$ that decrease with increasing pressure.
Data at higher pressures also have apparently linear regions in Fig. \ref{viscosity.rate.atT293K} with a slope near $-1$ that corresponds to $n$ near zero.
However, as noted already, the data at these pressures are better fit by the Eyring equation where the high frequency limit of Eq. \ref{eq:highrate.eyring.viscosity.rate},
$\eta \sim \log \dot{\gamma} / \dot{\gamma}$, can appear to be a power
law with $n$ near zero.
Eyring theory describes the decrease in viscosity by six orders of magnitude relative to $\eta_N$. 

The Eyring fits to simulation data for $P \geq 300$ MPa at high strain rates
($\dot{\gamma} > 10^6 \ \textrm{s}^{-1}$) give Newtonian viscosities very close to the experimental values.
As shown in Ref. \cite{jadhao.robbins.2017}, this approach successfully reproduces published experimental data for $\eta_{N}$ over a wide range of $T$, $P$, and $\eta_{N}$.
The only significant difference between Eyring fits and nonequilibrium experiments in Fig. \ref{viscosity.rate.atT293K}
occurs at the highest experimental rates where the experimental values drop slightly more rapidly than the fits.
The experiments used an ensemble that allowed dilation, which would reduce
the shear stress.
As discussed in Section \ref{sec:ensemble}, we expect this effect to be small.
Another possibility is that dissipation raises the experimental temperatures,
which also lowers the viscosity and must often be corrected for in EHL measurements \cite{spikes.tbl.2014,dini.spikes.2017}.

\begin{figure}[t] 
\centerline{\includegraphics[scale=0.37]{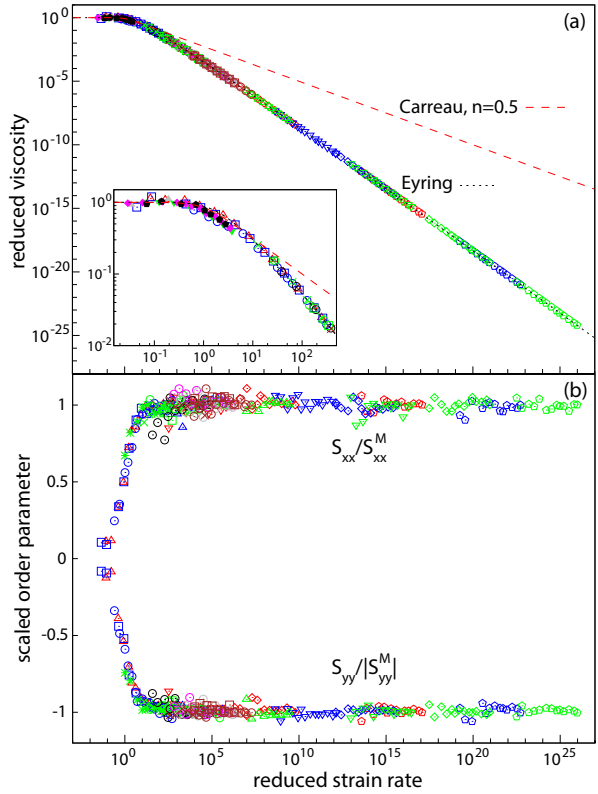}}
\caption{(a) Scaled viscosity vs. scaled rate for conditions where the Newtonian viscosity of squalane $\eta_{N} \gtrsim 1$ Pa-s.
All data follow the Eyring form (black dashed line).
Carreau theory with $n=1/2$ is also shown (red dashed line).
The inset shows a closeup of the initial shear thinning regime.
For clarity, only data points from experiments (filled symbols) and 
simulations (open symbols) at the same state points are compared to 
Eyring and Carreau fits.
(b) Order parameters $S_{xx}$ and $S_{yy}$ scaled by their maximum magnitudes at high rates.
Results for the order at all state points collapse when plotted against the same reduced rate that collapses viscosities.
Note that alignment saturates by a reduced strain rate of about 10 where the viscosity has dropped by less than an order of magnitude.
}
\label{master.high.viscosity.rate}
\end{figure}

As noted above, data for different temperatures and pressures are typically collapsed using the hypothesis of time-temperature superposition.
For systems that follow the Eyring equation, $\eta/\eta_{N}$ should follow
the universal curve of Eq. \ref{eq:eyring.viscosity.rate}
when the rate is scaled by $\dot{\gamma}_E$.
Figure \ref{master.high.viscosity.rate}(a) shows that this collapse applies for all systems with high Newtonian viscosity, $\eta_{N} \gtrsim 1$ Pa-s.
The collapse only requires a two parameter fit to a straight line
plot of $\sigma$ vs. $\log \dot{\gamma}$ at each $T$ and $P$.
Over almost 30 decades, simulation and experimental data lie on the analytic solution of the Eyring model.

The inset of Fig. \ref{master.high.viscosity.rate}(a) shows a closeup of the region where $\eta$ begins to decrease.
The range of $\eta/\eta_{N}$ (down to 0.01) is chosen to correspond to the range of reduced viscosity data shown in Fig. 1 of Ref. \cite{bair.cummings.prl.2002}.
The experimental data only extend down to $\eta/\eta_{N} \sim 0.4$. Over this limited experimental range, the Eyring equation (dotted line) is nearly the same as the Carreau equation with the value of $n \approx 1/2$ used in Ref.~\cite {bair.cummings.prl.2002} (dashed line).
To clearly distinguish the two forms requires a decrease in $\eta/\eta_{N}$ by a decade or more.

\begin{figure}[t] 
\centerline{\includegraphics[scale=0.37]{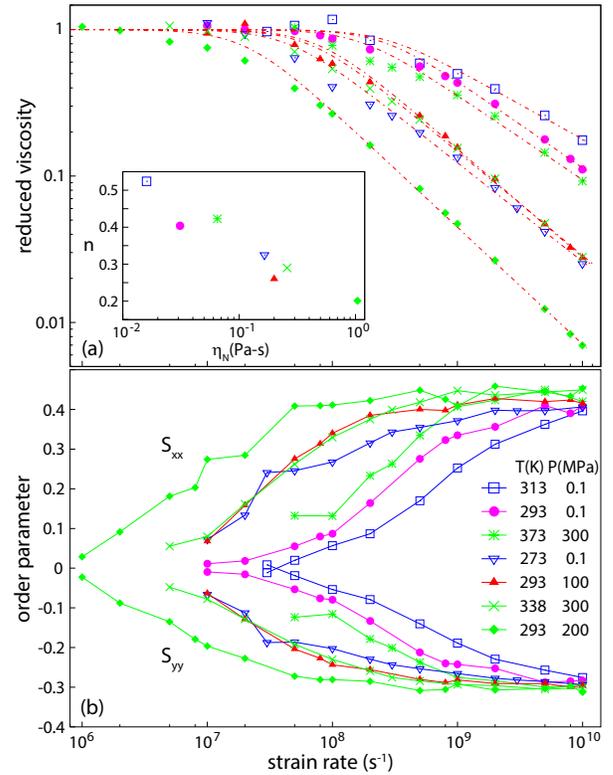}}
\caption{(a) 
	Steady-state shear viscosity normalized by $\eta_N$ vs. rate for squalane in conditions where $\eta_{N} \lesssim 1$ Pa-s. Symbols correspond to the values of $P$ and $T$ in the legend of (b). Red broken lines are best fits to the Carreau model. The corresponding values of $n$ are plotted against $\eta_N$ in the inset.
(b) Plots of the order parameters $S_{xx}$ (top) and $S_{yy}$ (bottom) as a function of rate.}
\label{master.low.viscosity.rate}
\end{figure}

The Carreau-like shear thinning of low viscosity fluids ($\eta_{N} \lesssim 1$ Pa-s) is
illustrated in Fig. \ref{master.low.viscosity.rate}(a), where
$\eta/\eta_{N}$ is plotted against the unscaled $\dot{\gamma}$.
Systems with larger $\eta_{N}$ begin to shear thin at lower rates and
show a larger total shear thinning that is fit by a higher $n$.
As shown in the inset of Fig. \ref{master.low.viscosity.rate}(a), $n$ correlates strongly with $\eta_N$, dropping from about 0.5 at the lowest viscosities to below 0.2 for the highest $\eta_N$.
Because of this change in $n$, the different curves do not collapse as well using time-temperature superposition as the data in Fig. \ref{master.high.viscosity.rate}(a).
As $\eta_N$ increases, the data move closer to being described by Eyring theory.
For cases like $P=200$ MPa and $T=293$ K with $n\leq 0.2$, Eyring theory provides a better fit as illustrated in Fig. \ref{atT293} where fits to both models are shown.

\subsection{Order parameters and shear thinning}

One commonly invoked mechanism for power-law shear thinning is a change in some order parameter.
For spherical molecules, this ordering can involve correlations between nearby molecules \cite{loose89,StevensRobbins1993}.
For squalane or chain molecules, the simplest type of order is alignment of molecules along the flow direction \cite{kroger1993,kroger1993b,barsky2002bulk,sivebaek2012effective}.
This may lower the viscosity by decreasing the rate of collisions between
molecules and the associated momentum transfer and dissipation.
In contrast, shear thinning in the Eyring model only involves a change
in the rate of rearrangements through activated motion over fixed barriers.

To study changes in order in squalane under shear we evaluated the
orientational tensor
\begin{equation}
S_{\alpha \beta}= \frac{3}{2}\left\langle \frac{1}{N_m} \sum_{i=1}^{N_m} u_{i\alpha}u_{i\beta} - \frac{1}{3}\delta_{\alpha \beta} \right\rangle, 
\end{equation}
where $\alpha$ and $\beta$ are Cartesian coordinates, $\hat{u}_i$ is the
unit vector in the direction between ends of the $i^{th}$ squalane molecule, and $N_m$ is the number of molecules.
The diagonal elements $S_{xx}$, $S_{yy}$, and $S_{zz}$ measure the degree of alignment of molecules
in the direction of the flow field, velocity gradient, and vorticity, respectively.
A value of 0 means that there is no preference to align along the axis,
while values of 1 and $-0.5$ imply perfect alignment along or perpendicular to the axis.

\begin{figure}[t]
\centerline{\includegraphics[scale=0.37]{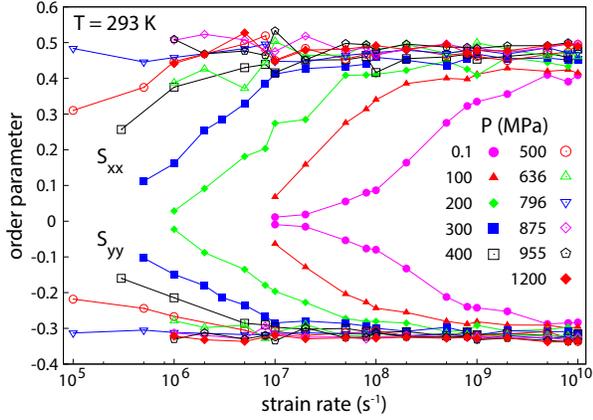}}
\caption{Rate dependence of $S_{xx}$ and $S_{yy}$ for squalane
at $T=293$ K and the indicated pressures.
}
\label{fig:OrderVsRate}
\end{figure}

\begin{figure}[t]
\centerline{\includegraphics[scale=0.37]{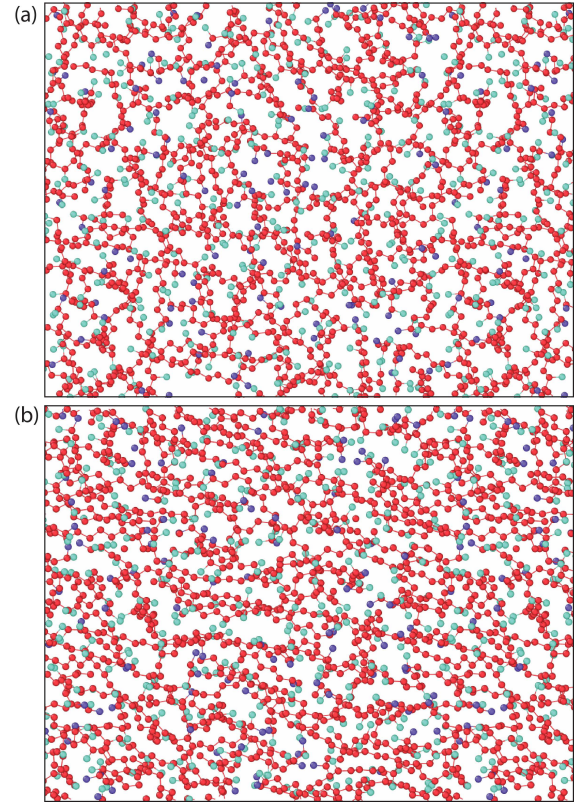}}
\caption{(a) Representative simulation snapshot of structure at $\dot{\gamma}=10^{7}$ $\mathrm{s}^{-1}$ for $P=0.1$ MPa and $T=293$ K. Here the order parameters $S_{xx}$ and $S_{yy}$ are nearly zero and no alignment is apparent in the snapshot. (b) Snapshot at the same rate and $T$ but at $796$ MPa. Here the order parameters have saturated and molecules show a clear tendency to align along the horizontal $x$-axis and away from the vertical $y$-axis.  In each case a narrow slice in the $x-y$ plane is shown for clarity. 
}
\label{fig:snapshot}
\end{figure}

Fig. \ref{fig:OrderVsRate} shows the variation of order with rate for
the $T=293$ K systems of Fig. \ref{atT293}.
Shear couples most strongly to the flow and gradient directions,
so only $S_{xx}$ and $S_{yy}$ are shown.
In equilibrium, the system must become isotropic with $S_{xx}=S_{yy}=S_{zz}=0$.
This limit is approached for the lowest pressures and rates in Fig. \ref{fig:OrderVsRate}.
With increasing rate, chains become more and more aligned along
the flow direction.
The value of $S_{xx}$ rises to about half the value for perfect order
and $S_{yy}$ drops to about $-0.3$.
The maximum magnitudes $S_{xx}^M$ and $|S_{yy}^M|$ rise only slightly with
increasing pressure and decreasing temperature.
Further increase in alignment appears frustrated by the antisymmetric
structure of the  squalane backbone that facilitates chain bending and suppresses crystallization.
Fig. \ref{fig:snapshot} (a) and (b) show snapshots of structure in the high rate limit for $P=0.1$ MPa and 796 MPa respectively at $T = 293$ K.
Note that a number of chains are bent and the structure retains the
disordered character of a glass.
Other molecules may adopt a more ordered structure that favors
shear localization.

For all $P$ and $T$, $S_{xx}$ and $S_{yy}$ rise rapidly  with rate and then saturate.
The rapid shear-thinning after order saturates must come from another mechanism.
To determine the amount of shear thinning that may come from alignment, we plot the reduced order parameter $S_{\alpha \alpha}/|S_{\alpha \alpha}^M|$
against reduced rate in Figs. \ref{master.high.viscosity.rate}(b) and
\ref{master.low.viscosity.rate}(b).
In the Eyring regime (Fig.~\ref{master.high.viscosity.rate}(b)), order saturates for reduced rates of about 10
where $\eta/\eta_N$ has only decreased by about a factor of 3.
In the Carreau regime (Fig.~\ref{master.low.viscosity.rate}(b)),
the rate at which order saturates and the value of $n$ decrease with increasing $\eta_N$, but order saturates by $\eta/\eta_N \sim 0.3$ for all cases.
The viscosity then continues to decrease at fixed order, dropping by a decade or more in several cases.

Estimates for how much shear thinning may come from alignment can be obtained from simulations of chain molecules.
Barsky and Robbins \cite{barsky2002bulk} considered the Kremer-Grest model for chains shorter than the entanglement length $\sim 85$ beads.
Power law shear thinning with $n \approx 0.5$ was observed for 16, 32, and 64 beads over up to 2 decades in rate and 1 decade in viscosity.
Chains continued to order over this range and the decrease in viscosity was directly correlated with a decrease in the extent of chains along the velocity gradient direction.

Baig {\it et al.} \cite{baig2010flow} have considered entangled chains in simple shear and found shear thinning by up to 3 orders of magnitude for chains
with 400 carbons. This is much longer than typical base oils in EHL fluids, but indicates that shear thinning due to alignment may extend over a wider range when molecular size increases.

We conclude that for squalane, molecules align along the flow direction, but the alignment can not be responsible for most of the shear thinning under EHL conditions.
Alignment saturates rapidly and the viscosity drops only by about a factor of three before saturation.
The viscosity then continues to drop by many decades with little
change in alignment.
The entire curve is well described by the Eyring model which does not rely on a change of structure, 
just a changing competition between shear rate and thermal activation.

\subsection{General features of shear thinning}

As noted above, bi-disperse Lennard-Jones (LJ) systems have been widely studied in research on the glass transition \cite{varnik2003shear,kob1995testing,rottler2005unified,varnik2006structural} and mechanical properties of amorphous metals \cite{rr.pre,falk1998dynamics,varnik04,maloney06pre,salerno13}.
Rottler and Robbins \cite{rr.pre} studied the rate-dependent stress
for the model described in Sec.~\ref{sec:LJglass} and found a logarithmic rate dependence consistent with Eyring theory for $T \leq 0.2 u/k_B$ and fairly Newtonian behavior for $T > 0.3 u/k_B$.
Here we examine the transition between these limiting behaviors.

\begin{figure}[t] 
\centerline{\includegraphics[scale=0.37]{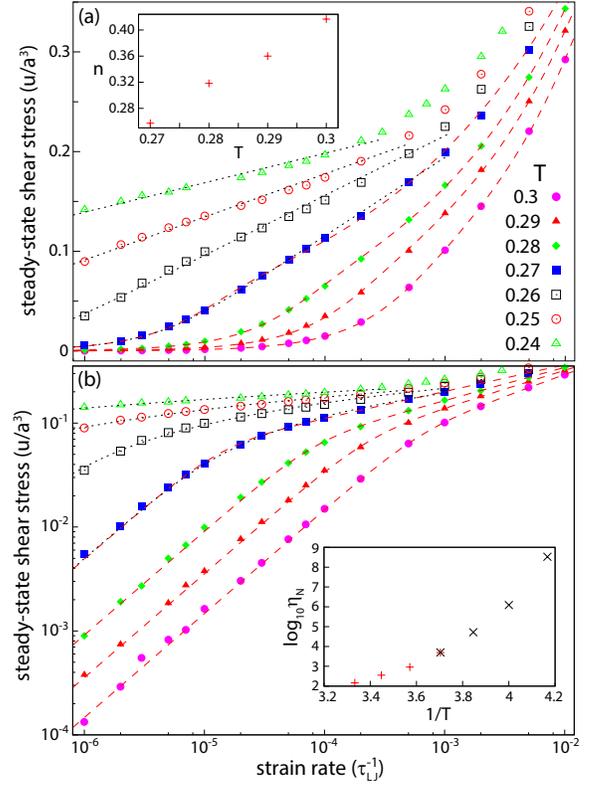}}
\caption{(a) Plot of stress vs. the logarithm of strain rate for the bi-disperse LJ system at the indicated values of $T$ in units of $u/k_B$.
	Red dashed lines show fits to the Carreau model at $T\geq 0.27$
	with values of $n$ shown in the inset.
	Black dotted lines show Eyring fits at $T \leq 0.27$.
	(b) Same data plotted on a log-log scale.
	The inset shows the variation of the logarithm of the Newtonian viscosity with inverse temperature.
Black crosses come from Eyring fits and red pluses from Carreau fits.
}
\label{LJglass}
\end{figure}

Figure \ref{LJglass} shows the rate dependent shear stress in the bi-disperse LJ system at temperatures between 0.24 and 0.3$u/k_B$.
This simple model system exhibits the same trends with temperature as squalane for stresses below 0.2$u/a^3$.
At high temperatures ($T\geq 0.28u/k_B$),
the stress can be fit to the Carreau equation with a power law $n$ that decreases from about 0.41 to 0.26 as $T$ decreases.
For $T \leq 0.27 u/k_B$, the curves follow the Eyring equation with an Eyring stress that decreases with $T$.

At low $T$ and stresses bigger than $0.2u/a^3$, the stress rises more rapidly
than predicted by Eyring theory.
This was also observed in Ref. \cite{rr.pre}.
Although the shear rate is low enough to prevent temperature increases,
the high disorder in the LJ system leads to non-affine deformations with large correlation lengths that relax very slowly \cite{tanguy}.
There is an increase in stress when the shear rate exceeds the relaxation
rate ($\sim 10^{-3} \tau_{LJ}^{-1}$).
While this regime may be of interest in the study of glassy systems, the rates are much larger than is relevant for EHL.

As part of the 10th Industrial Fluid Properties Simulation Challenge \cite{challenge2018}, Galvani and Robbins studied the small molecule
2,2,4-trimethylhexane with an all-atom potential over pressures from 0.1 MPa to 1 GPa and $T=293$ K \cite{Galvani2019}.
They also saw a crossover from Carreau behavior at low $\eta_N$ (low $P$)
to Eyring response at high $\eta_N$.
The low $P$ viscosity ($\sim 0.5\cdot 10^{-3}$ Pa-s) is smaller than squalane ($\sim 0.03$ Pa-s) and the transition to Eyring behavior occurs at a lower $\eta_N$.
The activation volume $V^* \sim 0.2$ nm$^3$ at room temperature is about 2/3 of the molecular volume.

A striking similarity between all three systems is that Newtonian behavior ends at almost the same shear stress in both Eyring and Carreau regimes.
For the Eyring model, $\eta$ drops to half of $\eta_{N}$ for $\sigma \sim 2\sigma_E$.
In Fig. \ref{atT293}, $\sigma_E \approx 10$ MPa and $\eta$ drops to about half $\eta_{N}$ when stress is between 10 MPa and 20 MPa.
In Fig. \ref{LJglass}, the Eyring stress at $T=0.27 u/k_B$ is
about $0.026 u/a^3$ and $\eta$ drops to half $\eta_{N}$ 
between one and two times this value.

For all three systems, the highest stresses reached at high rates are only an order of magnitude larger than the shear stresses at the upper end of Newtonian scaling.
We can estimate an upper bound for the high-rate stress using the physical picture underlying Eyring theory.
The energy barrier for activation vanishes for $\sigma _{max} = E/V^*$.
Eyring theory assumes that the stress is always smaller than this so that activation over a barrier dominates the flow rate.
The value of $\sigma_{max}$ can be measured using simulations at $T=0$ K where there is no thermal energy to activate transitions, and energy barriers to flow must vanish at the yield stress.
Simulations for squalane at 1 K give a yield stress of about 80 MPa at ambient pressure and
$\sigma_{max}$ rises linearly to about 250 MPa at 1.2 GPa \cite{Jclemunpub}.
As expected, these athermal yield stresses are comparable to but larger than the high rate stresses at the same pressure in Fig. \ref{atT293}.
Similarly, the low temperature yield stress for the bi-disperse LJ system, $\sim 0.55u/a^3$, is comparable to the high rate stresses in Fig. \ref{LJglass}.

Power law models like Carreau do not lead to an upper bound on stress in the same way.
To describe the slower than power law rise in $\sigma$ observed in some EHL fluids,
the Carreau model
has been modified to include a limiting stress as a separate parameter \cite{spikes.tbl.2014,Habchi2013}.
This saturation has been attributed to a change in flow mechanism
\cite{Bair2007,Habchi2010}, 
but without direct evidence in most cases.
This increases the number of fit parameters at each $T$ and $P$ to 4 compared to 2 for Eyring.

The limited range of stresses between the Newtonian and high-rate limits places bounds on the range of power law scaling. If the power law region of shear thinning covers less than a decade in shear stress, it covers less than $1/n$ decades in strain rate. For the common fit parameter of $n=1/2$, one should see
power law scaling over less than 2 decades in rate and one decade in viscosity. Fits to Carreau scaling that we have found in the EHL literature are
typically over smaller ranges
\cite{Bair2007,spikes.tbl.2014,bair.tbl.2015,Spikes2015,bair.cummings.prl.2002,Habchi2010}.
In contrast, the logarithmic increase in stress with rate in the Eyring equation means that it can fit over many decades in rate and the viscosity will
decrease by almost the same number of decades.
Hence we have almost 30 decades of scaling in Fig. \ref{master.high.viscosity.rate}(a) but only one or two in Fig. \ref{master.low.viscosity.rate}(a).

Past simulations of short chain molecules show the same trends we see for squalane and the bi-disperse LJ system.
Baljon and Robbins \cite{baljon1996energy} studied the Kremer-Grest model with 16 beads and saw a change from Eyring behavior at
$T=0.3u/k_B$ to power law scaling over two
decades in rate at $T=0.6 u/k_B$.
Earlier simulations of confined chains found similar
changes in behavior as $\eta_{N}$ increased due to decreasing system dimensions or increasing pressure \cite{thompson1992,thompson1995structure}.
These simulations \cite{thompson1992,thompson1995structure} found different exponents for shear-thinning
at constant volume, $n=1/2$, and constant pressure along the velocity gradient, $n=1/3$.
The value of $n=1/2$ is comparable to results shown above and in bulk studies of unconfined
chains \cite{barsky2002bulk}, which were also
at constant volume.
The rise in shear stress is slower when pressure is controlled because the system dilates to facilitate flow. This may happen in many experiments on EHL fluids as discussed in
Sec. \ref{sec:ensemble}.
At very high pressures, the value of $n$ was near unity. This was associated with a transition to glassy behavior \cite{thompson1995structure},
but is consistent with the above mentioned crossover to Eyring behavior found in squalane and the bi-disperse LJ system.

Sivebaek et al. \cite{sivebaek2012effective} considered confined hydrocarbon chains with 20 to 1400 carbons.
	They found power law scaling for $\dot{\gamma} =10^8$ to $3\cdot 10^{10} \ \textrm{s}^{-1}$
with varying $n$.
For $T>500$ K, 20 carbon chains had $n$ near unity, indicating Newtonian
behavior.
For $T < 200$ K, $n$ was close to zero and we find results in the Eyring regime give $n=0$ to $0.2$ when the viscosity is fit to a power law over two decades.
At intermediate $T$, they obtained $n=0.75$ and 0.3.
Longer chains also showed a similar transition at higher temperatures.
Their work was partly motivated by experiments \cite{yamada2002} that found
$n = 0.1\pm 0.1$ for a range of fluids. It seems likely that this experimental data could also be fit by Eyring theory.

\subsection{Variation of pressure and normal stress with rate in squalane}\label{sec:stresstensor}

\begin{figure}[t]
\centerline{\includegraphics[scale=0.5]{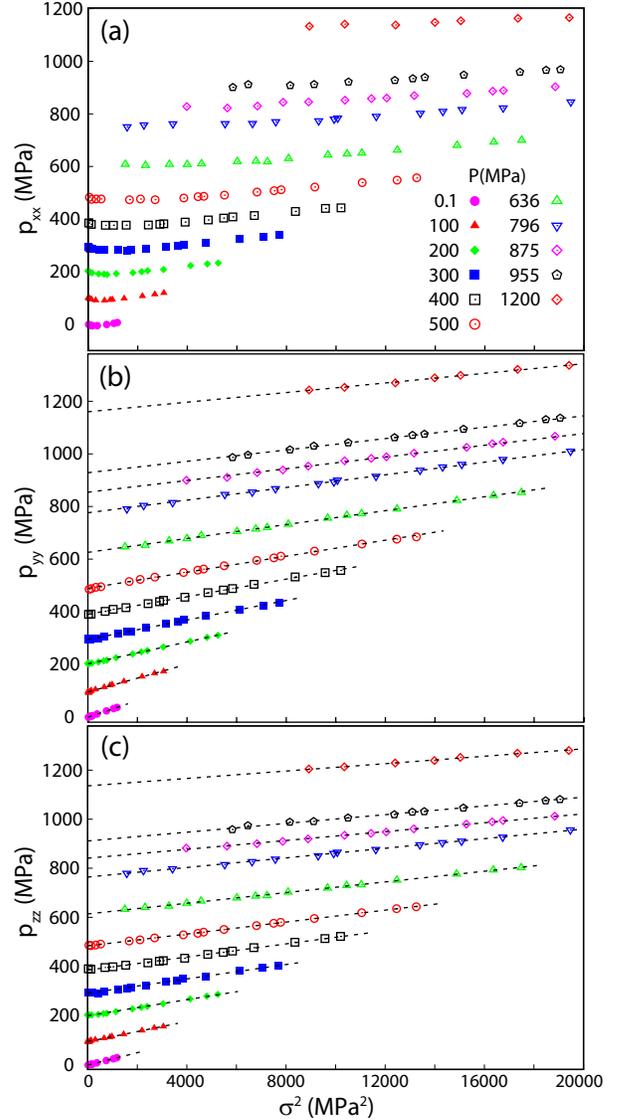}}
\caption{Diagonal components of the pressure tensor along (a) the flow direction $x$, (b) the gradient direction $y$, and (c) the vorticity direction $z$ plotted against the square of the shear stress, $\sigma$, at $T=293$ K and the equilibration pressures listed in (a).
	Dashed lines in (b) and (c) show linear fits to the data.
	Statistical errors are smaller than the symbol size.
}
\label{fig:pzzVSshearsq}
\end{figure}

In general, the stress tensor for simple shear has three different diagonal
components in addition to the shear stress studied above.
By convention, the pressure tensor has the opposite sign from the stress tensor
and the average of the diagonal components of the pressure tensor gives 
the pressure $P$.
The differences between the diagonal components are called normal stresses $N_i$
and only two are needed to completely specify the stress tensor.
These are usually taken to be
$N_1\equiv \sigma_{xx}-\sigma_{yy}=p_{yy}-p_{xx}$ and
$N_2\equiv \sigma_{yy}-\sigma_{zz}=p_{zz}-p_{yy}$,
where $x$ is the flow direction, $y$ is the velocity gradient direction, and $z$ is the vorticity direction.

Forcing the fluid to shear makes it hard for molecules to remain in a dense packing and favors dilation \cite{Reynolds1885}.
If the density is fixed, the pressure typically rises with rate.
In the Newtonian limit, symmetry arguments imply that the stress can only depend on even powers of the rate, and the $p_{ii}$ initially vary as $\dot{\gamma}^2\propto \sigma^2$.
The three diagonal components of pressure are plotted against shear stress squared in Fig. \ref{fig:pzzVSshearsq}. 
The component along the velocity, $p_{xx}$, drops initially with $\sigma^2$
and then rises slowly.
The change in slope occurs at stresses of $\sim 10$ to 20 MPa,
which is where the Newtonian regime ends in Fig. \ref{atT293}.
The two other diagonal components show a surprisingly linear dependence on $\sigma^2$ over the entire range of data.
Moreover, the prefactor changes relatively slowly with pressure at pressures of 200 MPa and above, where Eyring theory fits the shear stress.
Small deviations from quadratic scaling with $\sigma^2$ are revealed by examining $N_2=p_{zz}-p_{yy}$ as discussed below. First we consider the limiting pressure at low rates.

At sufficiently low pressures ($P \leq 500$ MPa at 293 K) and high temperatures, our simulations reach the Newtonian regime and all $p_{ii}$ converge to the equilibrium pressure at low rates. As noted above, this pressure may be slightly different than the pressure
used to create our initial states because of the long time for pressure equilibration in high viscosity fluids.
The differences are within our statistical fluctuations of about 5 MPa for $P \leq 300$ MPa but extrapolation gives 384 MPa and 483 MPa for the cases where our targets were 400 and 500 MPa, respectively.
At higher pressures, the dynamics are too slow for simulations to reach steady state, but quadratic fits in Fig. \ref{fig:pzzVSshearsq}(b) and (c) can be used to estimate the equilibrium pressure for a given density.
For example, extrapolating data for the more slowly changing $p_{zz}$ gives
614, 764, 841, 911, and 1135 MPa for target pressures of 636, 796, 875, 955, and 1200 MPa respectively. This corresponds to an offset of -3.3 to -5.4 \% with
errorbars on the extrapolated values of about 5 MPa.
The slopes of the quadratic fits drop with increasing pressure,
changing from about 0.0253 MPa$^{-1}$ at 0.1 MPa to 0.0076 MPa$^{-1}$ at 1200 MPa.
Values extrapolated from $p_{yy}$ are about midway between the values extrapolated from $p_{zz}$ and the target pressures; we obtain 625, 776, 854, 928, and 1160 MPa for the target pressures of 636, 796, 875, 955, and 1200 MPa respectively.

\begin{figure}[tb]
\centerline{\includegraphics[scale=0.36]{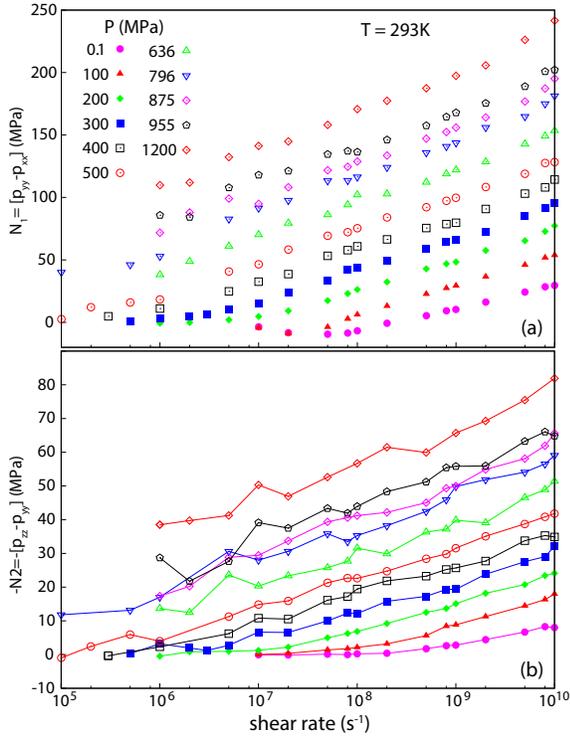}}
\caption{Normal stresses (a) $N_1$ and (b) $-N_2$ as a function of the logarithm of rate at $T=293$ K and the indicated target equilibration pressures.
	Error bars are of order 5 MPa which is larger than the symbol size. Lines are included to guide the eye in panel (b) because the statistical fluctuations are comparable to the spacing between data for adjacent pressures.
}
\label{fig:NVSrate}
\end{figure}

\begin{figure}[tb]
\centerline{\includegraphics[scale=0.34]{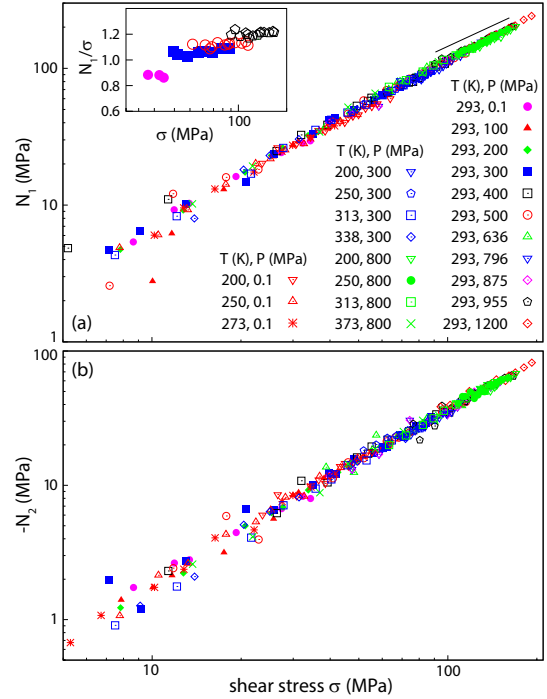}}
\caption{Normal stresses (a) $N_1$ and (b) $-N_2$ plotted
	against shear stress for the indicated values of $T$ and $P$.
	The inset in (a) shows high stress data for $N_1/\sigma$ at 
	$P=0.1$, 300, 500, and 955 MPa.
	A short solid line in (a) shows a slope of unity.
}
\label{fig:NVSrate2}
\end{figure}

Rheologists typically focus on the normal stresses that quantify differences between the diagonal components.
Figure \ref{fig:NVSrate} shows the variation of normal stresses with $\log \dot{\gamma}$ for $T=293$ K.
Except at very low rates in low viscosity fluids we find $p_{yy} > p_{zz} > p_{xx}$, so that $N_1$ gives the magnitude of the full spread between minimum
and maximum diagonal components and $N_2$ is negative and smaller in magnitude.
As noted above, the normal stresses should vary quadratically with rate
at low rates where $\sigma$ rises roughly linearly with rate.
From Fig. \ref{atT293}(b) we see that this nearly linear regime ends for all pressures when $\sigma \sim 10$ MPa.
The expected quadratic behavior in this limit is difficult to resolve because
$N_1$ and $|N_2|$ are only larger than statistical errors ($\sim 5$ MPa)
for a limited range of rates.
In the nonlinear regime where $\sigma > 10$ MPa, both $N_1$ and $|N_2|$ show a
clear logarithmic dependence on rate
that is similar to that predicted for $\sigma$ in Eyring theory.
Surprisingly, the logarithmic dependence in $N_1$ is seen clearly even for low pressure systems
where $\sigma$ is better fit by a power law Carreau form.

Figure \ref{fig:NVSrate2} shows that there is a strong connection between
the shear and normal stresses.
Results for $N_1$ and $|N_2|$ are plotted against $\sigma$ for a wide range of $T$ and $P$.
A good collapse is obtained for all state points at $\sigma \geq 20$ MPa.
This includes both low $\eta_N$ systems that follow Carreau scaling and high $\eta_N$ systems that follow Eyring. 
For $\sigma < 20$ MPa one enters the Newtonian regime (Fig. \ref{atT293}(b)) and the results for different systems 
begin to separate.
In this limit the normal stresses should scale as $\dot{\gamma}^2\propto \sigma^2$, but with prefactors that depend on $\eta_N$ and other quantities that
vary with $T$ and $P$, and our statistical uncertainties of a few MPa become
comparable to the normal stresses.

Closer analysis of the data in Fig. \ref{fig:NVSrate2}(a) shows that
for each curve, $N_1/\sigma$ is nearly constant at high stresses.
As shown for several $P$ at 293 K in the inset of Fig. \ref{fig:NVSrate2}(a), this ratio rises from about 0.9 at $P=0.1$ MPa to 1.2 at $P=955$ and 1200 MPa.
Similar linear behavior has been seen in past experiments and simulations \cite{bair.tbl.2015,bair.tribint.2004}, with $N_1/\sigma$ between 1 and 2.
Ref. \cite{bair.tbl.2015} argues that this scaling is inconsistent with Eyring theory, but it may be more accurate to say that it is not typically considered
in Eyring theory.
Fig. \ref{fig:NVSrate2} shows that fluids exhibit the same linear scaling of $N_1$ with $\sigma$ whether their shear thinning follows a Carreau or Eyring form.
Ref. \cite{bair.tbl.2015} also states that ``normal stresses conclusively demonstrate that alignment occurs and is responsible for shear thinning of simple liquids.''
We have already shown that shear thinning occurs after alignment saturates and
Fig. \ref{fig:NVSrate2} shows that results for fluids with varying degrees of alignment collapse on a common curve.

The ratio $\sigma/N_1$ is analogous to a friction coefficient.
Studies of the yield of glasses and granular systems also find
that the shear stress required for motion is proportional to the extra pressure along the direction perpendicular to the planes that slide over each other ($x-z$ planes here) \cite{Nadai1963,Schellart2000}.
For granular systems the friction coefficient is typically in the range 0.6 to 1 \cite{Schellart2000}, which is similar to the values inferred from the inset of Fig. \ref{fig:NVSrate2}(a) and past work \cite{bair.tribint.2004}.

In contrast to $N_1$, $|N_2|$ rises more rapidly than linearly with shear stress.
Given the limited range of scaling it is hard to determine an effective
exponent, but it is less than 2.
This implies that $p_{yy}$ and $p_{zz}$ can not both follow the linear behavior
in $\sigma^2$ that is suggested by Fig. \ref{fig:pzzVSshearsq}(b) and (c).
The magnitude of the difference in extrapolated zero-rate pressures from $p_{yy}$ and $p_{zz}$ ($\sim 10$ MPa) is comparable to the value of $N_2$ at the end of the Newtonian regime ($\sigma \sim 20$ MPa) where the scaling may be less universal.
Further studies on other systems and with higher statistics are needed to determine why quadratic scaling is observed and what its limitations are.

\begin{figure}[tb]
\centerline{\includegraphics[scale=0.34]{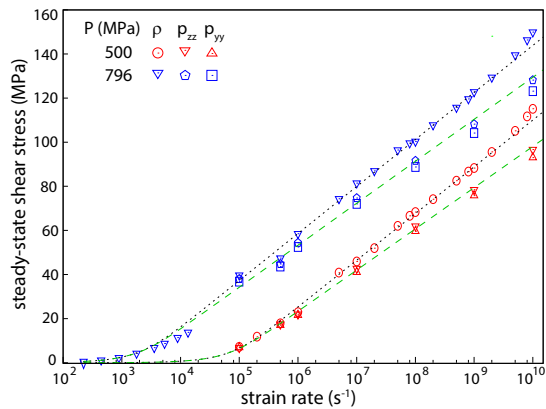}}
\caption{Shear stresses in squalane as a function of strain rate for systems
	in constant density $\rho$, constant $p_{zz}$, and constant $p_{yy}$ ensembles
	are indicated by points at $\dot{\gamma} \geq 10^5$ $\textrm{s}^{-1}$.
	Points at lower rate are from experiment at $p_{zz}=796$ MPa \cite{bair.cummings.prl.2002}.
	The equilibrium densities were obtained by equilibrating at 500 and 796 MPa,
	but the actual equilibrium pressures were determined to be 483 and 764 MPa.
	Data for constant $p_{zz}$ and $p_{yy}$ were obtained by interpolating our
	results to these equilibrium values.
	Black dotted lines show Eyring fits at constant density and green dashed lines are Eyring fits to constant $p_{zz}$ data. 
}
\label{fig:ensemble}
\end{figure}

\subsection{Shear thinning at constant pressure}\label{sec:ensemble}

The results presented in previous sections are all at constant density, while experiments may be performed in a different ensemble.
The nonequilibrium experiments shown in Fig. \ref{eyring.carreau} were done in
a Couette rheometer with approximately constant $p_{zz}$ \cite{bair.proc.inst.mech.2002}.
Measurements in EHL contacts are typically made with constant or measured $p_{yy}$ \cite{spikes.tbl.2014}.
In this section we discuss
how the choice of ensemble may affect the variation of $\sigma$ with $\dot{\gamma}$.

Figure \ref{fig:pzzVSshearsq} allows us to estimate the effect of dilation
in the constant $p_{zz}$ measurements shown in Fig. \ref{eyring.carreau}.
The quadratic fits in Fig. \ref{fig:pzzVSshearsq} imply that $p_{zz}$ would
only change by about 1 MPa at constant density for the largest
experimental stresses in Fig. \ref{eyring.carreau},
$\sigma \sim 10$ MPa.
Thus the difference in ensemble should not be significant in these experimental
conditions.
In our simulations or in experiments \cite{spikes.tbl.2014,dini.spikes.2017} that reach higher shear stresses,
constant density and constant $p_{zz}$ or $p_{yy}$ ensembles may give significantly different rate dependence.
For example, one can see from Fig. \ref{fig:pzzVSshearsq} that if $p_{zz}$ is fixed at 500 MPa, the density will drop from that associated with an equilibrium pressure of $500$ MPa to that associated with 400 MPa.
From Fig. \ref{eyring.carreau}, the stress at the highest rate would
drop from about 115 MPa to 100 MPa.

A full discussion of rate dependence at constant $p_{zz}$ and $p_{yy}$ will require new simulations at varying density, but we can generate approximate
curves by interpolation.
We focus on the Eyring regime where changes in density are larger and
uncertainties are small because changes with equilibrium pressure
at fixed rate are fairly linear.
Figure \ref{fig:ensemble} shows data for the densities corresponding to target equilibration pressures of 500 and 796 MPa.
The equilibrium pressures are estimated as 483 and 764 MPa and shear stresses interpolated to these values of $p_{zz}$ and $p_{yy}$ are also shown.
Both sets of constant pressure data show a linear dependence on the logarithm of rate.
Fits to Eyring theory for constant $\rho$ and $p_{zz}$ with the equilibrium $\eta_N$ at each target pressure are also shown.
The fits for different ensembles converge as the shear stress drops below $\sim 20$ MPa.
As expected, there is little difference at the low stresses of the rheology experiments, which are also shown for 796 MPa
\cite{bair.cummings.prl.2002}.
At high stresses the main difference between the ensembles is the slope, which corresponds to the Eyring stress.
The value of $\sigma_E$ drops from about 9.3 MPa at constant density to about 8.2 MPa at constant $p_{zz}$.
Constant $p_{yy}$ results are only slightly below constant $p_{zz}$ results, suggesting that rheometer
and model EHL contact experiments may give similar results even though they sample different ensembles.

\section{Discussion and Conclusion}

One of the challenges to resolving debates about the rheological response of fluids in EHL conditions is that the limited range of experimental data can often be fit with competing models or must be corrected for heating and other effects that are difficult to quantify.
The above results provide a comprehensive picture of shear thinning in
squalane as a function of temperature $T$ from 150 to 338 K and equilibrium pressures $P$ from
0.1 to 1200 MPa, corresponding to Newtonian viscosities $\eta_{N}$ from $10^{-2}$ Pa-s to more than $10^{12}$ Pa-s.
The range of rates studied overlaps with those in EHL contacts and extends up to
the rates where temperature becomes difficult to define,
$\dot{\gamma}_{max} \sim 10^{10}\, \textrm{s}^{-1}$ (Eq. \ref{eq:ratemax}).

Two different regimes are identified. At high $T$ and low $P$, where $\eta_N$
is low, the rheological response can be fit to the Carreau (power-law) model.
As $\eta_N$ increases, the rheological response changes and
becomes consistent with Eyring theory.
The transition from Carreau to Eyring occurs at a roughly constant value of $\eta_N$ for increasing
$P$ or decreasing $T$.
To test the generality of this transition, we also presented results for the
bi-disperse LJ system.
As for squalane, the rheological response in the bi-disperse LJ system follows the Carreau model when
$\eta_N$ is low and Eyring theory when $\eta_N$ is large.
The same transition is observed in all atom simulations of 2,2,4-trimethylhexane
that provided the best agreement with experimental $\eta_N$ in the 10th Industrial Fluid Properties Simulation
Challenge
\cite{Galvani2019}.

The transitions from Carreau to Eyring in squalane and the bi-disperse LJ system show several common features.
The first concerns time-temperature superposition, which assumes results from
different $T$ and $P$ can be
collapsed onto a universal rheological response function by a
simple rescaling with $\eta_N$ and a characteristic relaxation time.
In the Carreau regime, the viscosity follows power law shear-thinning
at high rates, but with a changing scaling exponent $n$.
For both systems, $n$ decreases from about 1/2
towards zero as the viscosity increases.
Because of this change in scaling behavior, data do not collapse well using
time-temperature superposition in the Carreau regime.
For systems where Carreau fits give $n$ less than about 0.2,
the data are better fit by Eyring theory.
The rheological response of all systems in the Eyring regime
obey time-temperature superposition.
For squalane,
rheological curves collapse
over more than 25 decades in viscosity and almost 30 decades in rate.

A second similarity between systems is that shear thinning occurs over
a relatively narrow range of stress.
Newtonian behavior extends up to a stress, $\sigma_N$, that is insensitive to $P$ and $T$
even though the rate required to reach this stress may change by orders of
magnitude.
For squalane, $\sigma_N \sim 10$ MPa, and for the
bi-disperse LJ system $\sigma_N \sim 0.03 u a^{-3}$.
The highest stresses reached in the shear thinning regime are only an order
of magnitude larger than $\sigma_N$.
Similar results are seen for 2,2,4-trimethylhexane.
The limited range of stress means that the power law Carreau model can only
show a decrease in viscosity of about an order of magnitude for $n=1/2$.
The drop increases to at most 2 orders of magnitude as $n$ decreases to
the lowest values in the Carreau regime.
In contrast, the decrease in viscosity in the Eyring regime is at least
2 orders of magnitude and extends to over 5 in our simulations for a given
$T$ and $P$ and to more than 25 orders of magnitude using time-temperature
superposition.

Finally, there are similar correlations between the viscosity $\eta_0$ at high $T$, low $P$ and high rates,  
and the viscosity at the transition between Carreau and Eyring.
In Fig. \ref{viscosity.rate.atT293K} one sees that all plots reach a viscosity $\eta_0 \sim 10^{-2}$ Pa-s at the highest rates, with a small rise as $P$ increases.
The Carreau/Eyring transition is at about 1 Pa-s, which is a factor of 100
larger than $\eta_0$.
The transition is at around 5000 $u \tau_{LJ}/a^3$ for the LJ system and the
high rate viscosity is once again about 2 orders of magnitude smaller.
A similar ratio is observed for 2,2,4 trimethylhexane
\cite{Galvani2019}.

It is natural that the Eyring model should be most accurate where $\eta_N/\eta_0$ is large, because it assumes the volumes being activated
vibrate many times in a fixed configuration before they pass over an
energy barrier.
This leads to a large value of the factor $e^{E/k_BT}$ in Eq. \ref{eq:newtonian.viscosity}.
For high $T$ and low $P$, where $\eta_N/\eta_0$ is small, this exponential factor is near unity, molecules are in nearly constant motion, the relaxation times are comparable to molecular vibrations,
and there is little shear thinning up to $\dot{\gamma}_{max}$.
Our results indicate that
Eyring theory describes the rheological response when $\eta_N/\eta_0$ has increased above 100,
implying that activation takes more than 100 vibrations.
Motion may still be dominated by activation
through discrete events at somewhat lower $\eta_N$,
but involve activation over a broad distribution of barrier heights.
As the time for activation increases, motion will be increasingly dominated by the lowest barrier and Eyring theory will provide a more accurate description.
It would be interesting to explore fits of the Carreau regime to a generalization of Eyring theory with multiple barriers \cite{ree1955theory} in future work. The presence of a single barrier controlling motion is also important to the scaling of $\eta_N$ with $T$. As noted in Ref. \cite{jadhao.robbins.2017}, $\eta_N$ is often fit to an Arrhenius form assuming a single energy barrier at high $T$ where $\eta_N$ is low. Our results for the nonlinear response show that a single barrier is not appropriate in this limit and should be more appropriate at low $T$.

Another mechanism of shear thinning is evolution of an order parameter (such as molecular alignment) as the shear rate becomes fast compared to the associated relaxation rate.
This can occur in a regime where local motion of atoms remains fast, and
successful theories have been developed for polymers \cite{Doi1988,larsonbook}.
These theories suggest that the range of shear thinning due to order should
be quite limited for small molecules like squalane and our simulations
bear this out.
The viscosity drops by only about a factor of three as the molecular alignment of squalane saturates to the high rate limit.
In the Carreau regime, there may be a causal relation between the alignment and shear thinning because local motion remains rapid.
In the Eyring regime, the direct contribution of alignment should be small.
Alignment is normally associated with entropic contributions that have
a characteristic scale set by $k_B T$.
In the Eyring regime, this becomes smaller than the energy barriers
that set the total stress.
A related problem is strain hardening of entangled glassy polymers.
In the melt, the stress can be calculated directly from the entropy associated with molecular alignment \cite{Doi1988,larsonbook}.
In the glass, strain hardening correlates with alignment but the
stresses are orders of magnitude larger than the alignment entropy
and are related to dissipation through activated rearrangements like those
assumed in Eyring theory \cite{hoy07,vanMelick03,kramer05}.

The above observations can be used to infer when EHL fluids should
be in different regimes.
The Carreau model is most likely to apply when the fluid operates at $T$ and $P$ where
$\eta_N$ is less than two orders of magnitude higher than $\eta_0$.
Given the power law nature of the Carreau model, this will typically be
at rates within 2-3 orders of magnitude of $\dot{\gamma}_{max}$.
Unless the molecules are large enough, this is likely to be at rates
at the high-rate edge of EHL conditions.
For squalane, shear thinning in the Carreau regime only occurs above $10^7$ $\textrm{s}^{-1}$ and for 2,2,4 trimethylhexane it starts at even higher rates.
Longer molecules, such as polymers, will have longer relaxation times that extend the Carreau regime to lower rates.
Shear-thinning due to alignment may also be relevant for long molecules and
may be estimated using existing rheological theories \cite{Doi1988,larsonbook}.
Eyring theory is most likely to apply in systems that have been pushed
towards the glassy regime by the high pressures in EHL contacts.
Given that enhanced resistance to squeeze out requires a substantial increase
in $\eta_N$, fluids in EHL contacts may often exceed the threshold of $\eta_N/\eta_0 \gtrsim 100$ needed to reach the Eyring regime.

In most applications, EHL fluids contain a mixture of molecules of different chemical structure and length.
There may be shear thinning associated with order of one species, say a long
polymer, that is dissolved in a lower viscosity base oil.
Carreau theory was originally developed for this case and has been successful
in fitting a range of experiments.
Future molecular simulations that provide detailed tests of its assumptions
would be valuable.
Simulations of glassy polymers suggest
that its validity will once again be confined to cases where the
viscosity of the base oil is not too much larger than its $\eta_0$.
For example, studies of alignment and shear hardening in mixtures of entangled and unentangled polymers show that the stress is related to dissipation due to activated hopping and obeys a simple mixing rule \cite{Hoy2009}.

Another important factor in comparing to experiments is that our simulations are at constant $T$ and $\rho$.
Experiments at EHL rates are prone to temperature increases because of the high rate of energy dissipation.
We have not considered this effect here, but methods of correcting for temperature rise have been used in recent simulations and experiments \cite{dini.spikes.2017}.
Experiments are also usually at a constant value of the pressure along the gradient or vorticity axis.
As discussed in Sec. \ref{sec:ensemble}, systems in the Eyring regime seem to continue to show an Eyring response, but with a reduced Eyring stress.
From the direct connection between pressure changes and shear stress shown
in Sec. \ref{sec:stresstensor}, we expect changes in rheological response to
be smaller in the Carreau regime where stresses are generally smaller.
Further work is needed in different controlled ensembles to quantify
effects more precisely.

Section \ref{sec:stresstensor} examined changes in the
diagonal components of the pressure tensor and the normal stresses.
There is a striking correlation between these quantities and the shear stress $\sigma$
that appears to be independent of whether the system is in the Carreau or Eyring regime.
In the shear-thinning regime, the diagonal components along the gradient and vorticity directions rise roughly quadratically with $\sigma$.
The normal stresses $N_1$ and $N_2$ for all $T$ and $P$ fall on universal curves when plotted against $\sigma$, independent of whether the shear thinning follows Carreau or Eyring theory. 
Moreover the ratio of $\sigma/N_1$ approaches a constant that is analogous to a friction coefficient and comparable to values for granular systems \cite{Nadai1963,Schellart2000}.
A similar proportionality between pressure and yield stress is seen in glassy
polymers, although with a lower value of $\sigma/N_1$ \cite{Rottler2001,quinson97}.
This proportionality has been seen before in EHL experiments and simulations
\cite{bair.tbl.2015,bair.tribint.2004}.
Our simulations show that it need not reflect alignment,
as had been inferred \cite{bair.tbl.2015}.
The value of $\sigma/N_1$ has been associated with shear localization,
which may be important in EHL performance \cite{dini.spikes.2017,bair.tribint.2004}.
Localization did not occur in our simulations, perhaps because of the small system size.
Simulations of larger systems, such as Ref. \cite{dini.spikes.2017} will be important to determine the onset of localization and the relation between
$\sigma/N_1$ and molecular structure.

\begin{acknowledgements}
We thank Scott Bair, Marco Galvani Cunha, and Hugh Spikes
for useful discussions.
This material is based upon work supported by the National Science
Foundation under Grant No. DMR-1411144. Financial support was also
provided by the Army Research Laboratory under the MEDE
Collaborative Research Alliance, through Grant W911NF-12-2-
0022. V.J. also thanks Indiana University Bloomington that supported this work through startup funds.
\end{acknowledgements}
\bibliography{References}   % name your BibTeX data base
\end{document}